\documentclass[aps,nofootinbib,prd,twocolumn,showkeys,10pt,floatfix]{revtex4-1}
\usepackage{graphicx}
\usepackage[retainorgcmds]{IEEEtrantools}
\usepackage{amssymb}
\usepackage{amsmath}
\usepackage[svgnames]{xcolor}
\usepackage{mathtools,slashed}
\usepackage{epstopdf}
\usepackage[utf8]{inputenc}
\usepackage{url}
\usepackage{subfiles}
\usepackage{physics}
\usepackage{tikz}
\usepackage[compat=1.1.0]{tikz-feynhand}
\usetikzlibrary{arrows.meta}
\usepackage{pgfplots}
\pgfplotsset{compat=1.18}
\usepackage{xparse}

\usetikzlibrary{positioning,arrows,patterns}
\usetikzlibrary{decorations.markings}
\usetikzlibrary{calc}
\usetikzlibrary {arrows.meta}

\usepackage{newtxtext}
\usepackage{newtxmath}
\usepackage[normalem]{ulem}
\usepackage[colorlinks,citecolor=DarkGreen,linkcolor=DarkRed,urlcolor=DarkBlue]{hyperref}

\usepackage{epsfig}

\newcommand{\be}{\begin{eqnarray}}
\newcommand{\ee}{\end{eqnarray}}
\begin{document}

\title{Excitation function for global \texorpdfstring{$\Lambda$}{Lambda} polarization in relativistic \\ heavy-ion collisions with the Core--Corona model
}

\author{Alejandro Ayala$^1$}
\author{Jos\'e Jorge Medina-Serna$^1$}
\email[Corresponding author:]{jose.medina@correo.nucleares.unam.mx}
\author{Isabel Dom\'\i nguez$^2$}
\author{Mar\'\i a Elena Tejeda-Yeomans$^3$}
\affiliation{$^1$Instituto de Ciencias
Nucleares, Universidad Nacional Aut\'onoma de M\'exico, Apartado
Postal 70-543, CdMx 04510,
Mexico}
\affiliation{$^2$Facultad de Ciencias F\'\i sico-Matem\'aticas, Universidad Aut\'onoma de Sinaloa, Avenida de las
Am\'ericas y Boulevard Universitarios, Ciudad Universitaria, Culiac\'an, 80000, Mexico}
\affiliation{$^3$Facultad de Ciencias-CUICBAS, Universidad de Colima, Bernal D\'\i az del Castillo No. 340,
Colonia Villas San Sebasti\'an, Colima, 28045, Mexico}

\begin{abstract}

We compute the excitation function of the global $\Lambda$ polarization in semicentral heavy-ion collisions within a Core--Corona framework, where the interaction region is described as a dense core and a dilute corona separated by a critical value of the participant density. An important ingredient in the model are the intrinsic polarization functions in each of the two regions. These are computed from a field-theoretical approach where the vortical motion of the medium is included in an effective fermion propagator, which we derive explicitly. The interactions in the core and the corona are transmitted by suitable mediators at finite temperature and baryon chemical potential; gluons for the former and $\sigma$-mesons for the latter. The temperatures and baryon chemical potentials are related to the collision energies along the chemical freeze-out curve. By allowing the cross section for $\Lambda$ production in the nuclear environment to take on values below the nucleon-nucleon threshold cross section, the calculation describes the lowest energy polarization data point. For the centralities corresponding to the experimental data, we find that the contribution from the corona is the dominant one and that a lifetime, and correspondingly a volume of this region, which becomes larger for the smaller energies due to stopping, is an essential ingredient in the calculation. Overall, the model provides a good description of the excitation function across the full experimental range and predicts a robust maximum near $\sqrt{s_{NN}}\sim$ 3 GeV that remains stable under reasonable
variations of the freeze-out curve and the proton-proton $\Lambda$ production threshold to account for subthreshold production in a nuclear environment.

\end{abstract}


\maketitle

\section{Introduction}

Heavy-ion collisions offer a unique opportunity to study, in a laboratory controlled environment, the properties of strongly interacting matter under extreme conditions of temperature and density. The knobs are provided by the center-of-mass collision energy and the centrality selection. Among the several outstanding properties of the matter produced in these reactions are the features of the polarization excitation functions of $\Lambda$ and $\overline{\Lambda}$ hyperons for semi-central collisions. Current heavy-ion experiments have observed that the global $\Lambda$ and $\overline{\Lambda}$ polarization excitation functions increase as  $\sqrt{s_{NN}}$ decreases~\cite{Gou:2024yrx,HADES:2022enx,STAR:2021beb,ALICE:2019onw,STAR:2018gyt,STAR:2025}. In the near future, the Multi-Purpose Detector (MPD)~\cite{MPD:2022qhn,MPD:2025jzd}, at the Nuclotron-based Ion Collider fAcility (NICA) and the Compressed Baryonic Matter (CBM) experiment at GSI-FAIR~\cite{Messchendorp:2025men}, will provide additional data precisely in the region where these functions increase~\cite{Tsegelnik:2024ruh,Troshin:2024nig,Nazarova:2024jic,Nazarova:2021lry,Ayala:2020vvs}.  

The possibility that the $\Lambda$ and $\overline{\Lambda}$  polarization is produced by the intense vortical motion generated in the interaction region of semi-central heavy-ion collisions has been extensively examined~\cite{Becattini2008,Becattini:2015ska,Becattini:2014yxa,Becattini2017,Huang:2020xyr,Sass:2022ucj,Karpenko:2021wdm}. The well-known Barnett effect, by which a spinning ferromagnet experiences a change in its magnetization~\cite{1915PhRv....6..239B} and the related Einstein–de Haas effect, by which a change in the magnetic moment of a free body causes the body to rotate~\cite{1915KNAB...18..696E}, support this expectation. Some consequences of this vortical motion in heavy-ion reactions for the restoration of chiral symmetry have been studied in Refs.~\cite{Gaspar:2023nqk,Hernandez:2024nev}.

Motivated by the observation of global $\Lambda$ and $\overline{\Lambda}$ polarization in heavy-ion collisions, a large amount of research has been carried out to look for the mechanisms that transfer the generated angular momentum into hyperon polarization. Several Monte Carlo and phenomenological approaches have been used: generators, such as the Multiphase Transport model (AMPT) containing hydro evolution, have been used to compute the global hyperon polarization in the collision energy range $\sqrt{s_{NN}}=1$ GeV -- 2.76 TeV~\cite{Guo:2022cxa,Wu:2020yiz,Wei:2018zfb,Xia:2018tes}. In the energy range $\sqrt{s_{NN}}=$ 3 - 200 GeV the Ultra-relativistic Quantum Molecular Dynamics (UrQMD) model combined with viscous hydro, the Heavy Ion Jet Interaction Generator (HIJING) and the microscopic transport model PACIAE, have also been used~\cite{Deng:2021miw,Deng:2020ygd,Karpenko:2016jyx,Lei:2021mvp,Deng:2016gyh}. Phenomenological analyses have focused on clarifying the microscopic origin of the transfer of rotational motion to spin. This transfer can only happen provided that the reaction induced by the medium occurs fast enough such that the alignment of the spin and the angular velocity takes place within the lifetime of the medium~\cite{Montenegro:2018bcf,Kapusta:2019ktm,Kapusta:2019sad,Montenegro:2020paq,Kapusta:2020dco,Kapusta:2020npk,Torrieri:2022ogj}. 

A great deal of effort has been devoted to understanding the microscopic mechanism that transfers vorticity to polarization in a heavy-ion reaction. Since the interaction region can be thought of as consisting of a dense enough subregion to produce QGP and a not so high density subregion where ordinary nuclear matter undergoes reactions, the modeling of the transfer of vorticity to spin degrees of freedom needs to account for particles emitted from these two subregions. This approach has been followed in recent works where the interactions in the QGP are mediated by gluons~\cite{Ayala:2021xrn,Ayala:2023xyn,Ayala:2022yyx,Ayala:2023vgv,Ayala:2019iin,Ayala:2023vgv,Ayala:2020ndx,Ayala:2020soy} and in the nuclear environment by $\sigma$-mesons~\cite{Ayala:2025bzk}. In both cases, the transfer of angular momentum to spin is accomplished by using a suitable fermion propagator, corresponding to quarks in the QGP and to protons in the nuclear medium, that contains the information of the rotating environment~\cite{Ayala:2021osy}. In this work, we compute the excitation function of the $\Lambda$ polarization within the Core--Corona model, accounting for the contributions from both the core and the corona subregions.

The work is organized as follows. In Sec.~\ref{secII} we present the general framework for computing the intrinsic $\Lambda$ polarization within a field-theoretical perspective, where the effects of the rotating environment are incorporated through an effective fermion propagator. In Sec.~\ref{secIII} we undertake the explicit computation of such propagator providing details of its derivation. In Sec.~\ref{secIV} we use this propagator to compute the interaction rate for the alignment of the $\Lambda$-spin and vorticity in the corona region at finite temperature and baryon density. We model the interaction between nucleons and $\Lambda$ as mediated by a $\sigma$-meson exchange using for that purpose the $\sigma$-meson propagator at finite temperature and baryon density recently derived in Ref.~\cite{Ayala:2025bzk}. In Sec.~\ref{secV} we combine all the ingredients to find the excitation function for the global $\Lambda$ polarization. We finally summarize and conclude in Sec.~\ref{concl}.

\section{Core-Corona model for \texorpdfstring{$\Lambda$}{Lambda} polarization}\label{secII}

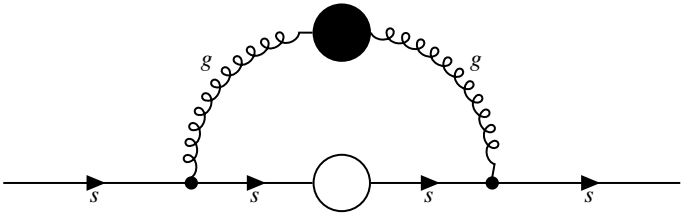
\begin{figure}[b]
    \centering
    \begin{tikzpicture}
			\begin{feynhand}
				\vertex (a) at (0,0); \vertex [dot] (b) at (2.5,0) {}; \vertex [blob] (c) at (4.5,2) {}; \vertex [dot] (d) at (6.5,0) {}; \vertex (e) at (9,0); \vertex [ringblob] (f) at (4.5,0) {};
				\propag [fer] (a) to [edge label' =$s$] (b);
				\propag [fer] (b) to [edge label' =$s$] (f);
				\propag [fer] (f) to [edge label' = $s$] (d);
				\propag [fer] (d) to [edge label' =$s$] (e) ;
				\propag [glu] (b) to [quarter left, edge label =$g$] (c);
				\propag [glu] (c) to [quarter left, edge label=$g$] (d);
			\end{feynhand}
\end{tikzpicture}
\caption{One-loop quark self-energy diagram describing the rate of spin-vorticity alignment for strange quarks in the QGP. The gluon line with a blob represents the effective gluon propagator at finite baryon density and temperature. The open circle on the strange quark propagator represents the effect of the rotating environment.}
\label{loopquark}
\end{figure}

In the Core--Corona model, the overlap region of the two colliding ions in peripheral heavy-ion collisions is decomposed into a central core and an outer corona. The core is modeled as a region with a sufficiently high density of participants to produce a thermalized quark-gluon plasma (QGP) fireball, whereas interactions in the corona resemble those in proton-proton collisions. As a result, particles originating from these two regions exhibit distinct features associated with different hadronization mechanisms, namely coalescence in the QGP and recombination (REC) processes in the corona. These features reflect the different interactions with the surrounding environment. We therefore expect observables such as polarization to depend not only on the collision energy and centrality, but also on the relative abundances of particles emitted from the core and the corona.

If the total number of $\Lambda$s coming from the core is $N_{QGP}^{\Lambda}$ and $N_{REC}^{\Lambda}$ is the number of $\Lambda$s coming from the corona, the Core-Corona model allows to express the global polarization as
\begin{eqnarray}
    \mathcal{P}^\Lambda &=& \frac{(N_{\Lambda\;QGP}^\uparrow+N_{\Lambda\;REC}^\uparrow) - (N_{\Lambda\;QGP}^\downarrow+N_{\Lambda\;REC}^\downarrow)}{(N_{\Lambda\;QGP}^\uparrow+N_{\Lambda\;REC}^\uparrow) + (N_{\Lambda\;QGP}^\downarrow+N_{\Lambda\;REC}^\downarrow)}\nonumber\\
    &=&\frac{(N_{\Lambda\;QGP}^\uparrow-N_{\Lambda\;QGP}^\downarrow) + (N_{\Lambda\;REC}^\uparrow-N_{\Lambda\;REC}^\downarrow)}{(N_{\Lambda\;QGP}^\uparrow+N_{\Lambda\;REC}^\uparrow) + (N_{\Lambda\;QGP}^\downarrow+N_{\Lambda\;REC}^\downarrow)}.
    \end{eqnarray}
   We define the intrinsic polarization of the core as
    \begin{eqnarray}
        z_{QGP}=\frac{N_{\Lambda\;QGP}^\uparrow-N_{\Lambda\;QGP}^\downarrow}{N_{\Lambda\;QGP}},
\end{eqnarray}
where $N_{\Lambda\;QGP}=N_{\Lambda\;QGP}^\uparrow+N_{\Lambda\;QGP}^\downarrow$ is the total number of $\Lambda$s coming from the core. In the same manner, we define the total number of $\Lambda$'s coming from the corona as $N_{\Lambda\;REC}=N_{\Lambda\;REC}^\uparrow+N_{\Lambda\;REC}^\downarrow$. Then, we define the intrinsic polarization of the corona as 
\begin{eqnarray}
    z_{REC}=\frac{N_{\Lambda\;REC}^\uparrow-N_{\Lambda\;REC}^\downarrow}{N_{\Lambda\;REC}}.
\end{eqnarray}
Therefore, 
\begin{eqnarray}
\mathcal{P}^\Lambda &=& \frac{N_{\Lambda\;QGP}}{N_{\Lambda\;QGP}+N_{\Lambda\;REC}}z_{QGP} + \frac{N_{\Lambda\;REC}}{N_{\Lambda\;QGP}+N_{\Lambda\;REC}}z_{REC}\nonumber\\
\nonumber\\
&=&\frac{\frac{N_{\Lambda\;QGP}}{N_{\Lambda\;REC}}}{1+\frac{N_{\Lambda\;QGP}}{N_{\Lambda\;REC}}}z_{QGP} + \frac{1}{1+\frac{N_{\Lambda\;QGP}}{N_{\Lambda\;REC}}}z_{REC}
\label{numbers}.
\end{eqnarray}

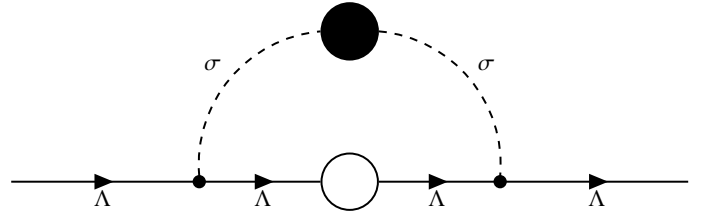
\begin{figure}[b]
    \centering
    \begin{tikzpicture}
			\begin{feynhand}
\vertex (a) at (0,0); 
\vertex  [dot] (b) at (2.5,0) {}; 
\vertex [blob] (c) at (4.5,2) {}; 
\vertex  [dot](d) at (6.5,0) {};
\vertex (e) at (9,0); 
\vertex [ringblob] (f) at (4.5,0) {};
				
                \propag [fer] (a) to [edge label' =$\Lambda$] (b);
				\propag [fer] (b) to [edge label' =$\Lambda$] (f);
				\propag [fer] (f) to [edge label' =$\Lambda$] (d);
				\propag [fer] (d) to [edge label' =$\Lambda$] (e) ;
				\propag [sca] (b) to [quarter left,  edge label =$\sigma$] (c);
				\propag [sca] (c) to [quarter left,  edge label=$\sigma$] (d);
			\end{feynhand}
	\end{tikzpicture}
\caption{One-loop $\Lambda$ self-energy diagram describing the rate of spin-vorticity alignment for $\Lambda$s in the corona region. The dashed line with a blob represents the effective $\sigma$ propagator at finite baryon density and temperature~\cite{Ayala:2025bzk}. The open circle on the $\Lambda$ propagator represents the effect of the rotating environment on the nucleons.}
\label{loop}
\end{figure}
Calculating the intrinsic polarization requires computing the relaxation times $\tau_{QGP}$ and $\tau_{REC}$, namely the times required for the spin to align with the vorticity in each of the two regions that make up the overall interaction region. In the core, $\tau_{QGP}$ was calculated in Ref.~\cite{Ayala:2023vgv} as a function of the collision energy and the impact parameter and used to find the $\Lambda$ and $\overline{\Lambda}$ polarization excitation functions, assuming that the contribution from the corona region was negligible~\cite{Ayala:2021xrn} and that the polarization of the strange quark is translated into the corresponding $\Lambda$ polarization during the hadronization process. The intrinsic polarization is expressed in terms of $\tau_{QGP}$ as a function of the QGP lifetime $\Delta\tau_{QGP}$, that is,
\begin{eqnarray}
    z_{QGP} = 1 - e^{-\Delta\tau_{QGP}/\tau_{QGP}}.
\end{eqnarray} 
The relaxation time is the inverse of the alignment rate, $\tau_{QGP} \equiv 1/\Gamma_{QGP}$, which is, in turn, obtained from the imaginary part of the self-energy $\Sigma^\pm_{QGP}$, depicted in Fig.~\ref{loopquark}. At one-loop order, the effects of the rotating environment are encoded in the loop quark propagator. As described in~\cite{Ayala:2023vgv}, the interaction rate for the alignment (antialignment) in the QGP for a strange quark with four-momentum $P=(p_0,\Vec{p})$ is given by
\begin{eqnarray}
    \Gamma^\pm_{QGP}(p_0)= \tilde{f}(p_0 -\mu_B \mp \Omega / 2)\text{Tr}\left[\text{Im}\Sigma^\pm_{QGP}\right],
\label{gammadefQGP}
\end{eqnarray}
where $\tilde{f}(p_0)$ is the Fermi–Dirac distribution, $\mu_B$ is the chemical potential of the quark and $\Omega$ is the angular velocity of the environment. The total alignment rate in the QGP is given by
 \begin{equation}
    \Gamma_{QGP}=V\int\frac{d^3p}{(2\pi)^3}\left[\Gamma^+_{QGP}(p_0)-\Gamma^-_{QGP}(p_0)\right],
    \label{gammaTot}
\end{equation}
where $V$ is the volume of the collision region.

Similarly, the intrinsic polarization for $\Lambda$s produced in the corona region is given by
\begin{eqnarray}
    z_{REC} = 1 - e^{-\Delta\tau_{REC}/\tau_{REC}},
\end{eqnarray} 
where, $\Delta\tau_{REC}$ is the lifetime of the corona. In analogy with the calculation of the relaxation time in the QGP, the relaxation time $\tau_{REC}$ can be computed from the $\Lambda$ self-energy $\Sigma^\pm_{REC}$, depicted in Fig.~\ref{loop}. 

To model the interactions that produce $\Lambda$s in the corona, which is mainly populated by nucleons, one can use an effective Lagrangian describing the $\Lambda$ interactions with mesons. We resort to the relativistic
mean-field (RMF) framework and consider that among the light mesons that couple to strange baryons, the main contribution comes from the isoscalar-scalar
$\sigma$-meson describing the attractive part of the hyperon-nucleon interaction
at low energies, which is, at the same time, the lightest degree of freedom contributing to the in-medium $\Lambda$ self-energy ~\cite{Li:2007zzh,Dutra:2014qga,Liu:2018img}. The Lagrangian density for $\Lambda$ hyperons that
interact with the scalar $\sigma$ is given by
\begin{equation}
\mathcal{L}_{\Lambda}=\bar{\psi}_\Lambda \left[ i\slashed{\partial} 
- M_\Lambda - g_{\Lambda}\sigma \right]{\psi}_\Lambda,
\label{eq:Lag}
\end{equation}
where $g_{\Lambda}$ is the $\Lambda$-$\sigma$ coupling and $M_\Lambda$ is the $\Lambda$ mass. While full RMF models for 
hypernuclear matter typically include the $\omega$ and $\rho$ vector mesons to account for repulsive interactions and isospin dependence, these contributions are not expected to dominate the spin-alignment mechanism
under consideration in a heavy-ion reaction given their larger mass. However, we notice that the same description is usually employed to model the interactions of hyperons in the
cores of neutron stars, where these particles play a central role in what is known as the
\lq\lq hyperon puzzle'' \cite{Oertel:2014qza,sun2018massive,german2022lambda}.

Before proceeding to the calculation of $\Sigma^\pm_{REC}$ and thus of $\tau_{REC}$ we pause to find the propagator for a fermion subject to the effects of a rotating environment. The calculation of this propagator corrects previous derivations found in Refs.~\cite{Ayala:2021osy, Ayala:2023vgv}.

\section{Propagator for a spin one-half fermion in a rotating environment}\label{secIII}

The physics in a relativistic rotating frame is most easily described in terms of an effective metric tensor resembling that of curved spacetime. For simplicity, we model the interaction region as a rigid cylinder rotating around the $\hat{z}$ axis with constant angular velocity $\Omega$, as expected in non-central heavy-ion collisions at early times. We can write the metric tensor as
\begin{equation}
  g_{\mu\nu}=  \begin{pmatrix}
1-(x^{2}+y^{2})\Omega^{2} & y\Omega & -x\Omega & 0\\
y\Omega & -1 & 0& 0\\
-x\Omega & 0&-1 &  0\\
0 & 0 & 0& -1\\
\end{pmatrix}.
\end{equation}
A fermion with mass $m$ within the cylinder is described by
the Dirac equation
\begin{equation}
    \left[i\gamma^{\mu}\left(\partial_{\mu} + \Gamma_{\mu} \right) -m \right]\Psi=0,
    \label{DiracEq}
\end{equation}
where $\Gamma_{\mu}$ is the affine connection. In this context, the $\gamma^{\mu}$-matrices in Eq.~(\ref{DiracEq}) correspond to the Dirac matrices in the rotating frame, which satisfy the usual anti-commutation relations 
\begin{equation}
    \{ \gamma^{\mu} , \gamma^{\nu}\}=2g^{\mu\nu}.
\end{equation}
The relation between the gamma matrices in the rotating frame and the usual gamma matrices is

\begin{equation}
    \begin{aligned}
        &\gamma^{t}=\gamma^{0}, \;\;\;\;\;\;\; \gamma^{x}=\gamma^{1}+y\Omega\gamma^{0},\\
        &\gamma^{z}=\gamma^{3}, \;\;\;\;\;\;\; \gamma^{y}=\gamma^{2}-x\Omega\gamma^{0}.
    \end{aligned}
\end{equation}
In this notation, $\mu=\{t,x,y,z\}$ refers to the rotating frame while $\mu=\{0,1,2,3\}$ refers to the local rest frame. Therefore, Eq.~(\ref{DiracEq}) can be written as
\begin{equation}
\begin{aligned}
      \Big[i\gamma^{0} & \left( \partial_{t}-x\Omega\partial_{y} + y\Omega\partial_{x}-\frac{i}{2}\Omega\sigma^{12} \right)\\ &
      +i\gamma^{1}\partial_{x}+i\gamma^{2}\partial_{y}+i\gamma^{3}\partial_{z} -m \Big]\Psi=0. 
\end{aligned}
\label{DiracEq2}
\end{equation}
In the Dirac representation, 
\begin{equation}
  \sigma^{12}=  \begin{pmatrix}
\sigma^{3} & 0\\
0 & \sigma^{3}
\end{pmatrix},
\end{equation}
where $\sigma^{3}=\mbox{diag}(1,-1)$ is the Pauli matrix associated with the third component of the spin. Therefore, we can rewrite Eq.~(\ref{DiracEq2}) as

\begin{equation}
\begin{aligned}
    \left[i\gamma^{0}\left(\partial_{t}+ \Omega\hat{J}_{z} \right) +i\vec{\gamma}\cdot\vec{\nabla}-m \right]\Psi=0,
\end{aligned}
\label{DiracEq3}
\end{equation}
where
\begin{equation}
    \hat{J}_{z}\equiv\hat{L}_{z}+\hat{S}_{z}=-i(x\partial_{y} - y\partial_{x})+\frac{1}{2}\sigma^{12}.
\end{equation}
This expression defines the total angular momentum in the $\hat{z}$ direction. The term $\hat{L}_{z}$ represents the orbital angular momentum, whereas $\hat{S}_{z}$ is the spin. On the other hand, the term  $-i\vec{\nabla}$ is the usual momentum operator. We can find solutions to Eq.~(\ref{DiracEq3}) in the form
\begin{equation}
    \Psi(x)= \left[i\gamma^{0}\left(\partial_{t}+ \Omega\hat{J}_{z} \right) +i\vec{\gamma}\cdot\vec{\nabla}+m \right]\phi(x),
    \label{psi}
\end{equation}
and then, the function $\phi(x)$ satisfies a Klein-Gordon like equation
\begin{equation}
    \left[\left(i\partial_{t}+\Omega\hat{J}_{z} \right)^{2} +\partial^{2}_{x}+\partial^{2}_{y}+\partial^{2}_{z} -m^{2}\right]\phi(x)=0.
    \label{KG}
\end{equation}
Notice that the spin operator $\hat{S}_{z}$, when applied to $\phi(x)$, produces eigenvalues $s=\pm$1/2. Consequently, conservation of the total angular momentum expressed in terms of the eigenvalues $j=s+l$ imposes solutions with $l$ for $s=$1/2 and $l+1$ for $s=-$1/2. With these considerations, the solution of Eq.~(\ref{KG}) can be written in cylindrical coordinates $(t, x, y, z) \to (t, \rho \sin\varphi, \rho \cos\varphi, z)$ as
\begin{equation}
    \phi^\lambda(x)=\begin{pmatrix}
J_{l}(k_{\perp}\rho) \\
J_{l+1}(k_{\perp}\rho)e^{i\varphi} \\
J_{l}(k_{\perp}\rho)\\
J_{l+1}(k_{\perp}\rho)e^{i\varphi}
\end{pmatrix}
e^{-Et+ik_{z}z+il\varphi},
\label{eigenbasis}
\end{equation}
where $J_{l}$ are Bessel functions of the first kind,
\begin{eqnarray}
k_{\perp}^{2}=\tilde{E}^{2}-k_{z}^{2}-m^{2},
\label{onmassshell}
\end{eqnarray}
is the transverse momentum squared and we have defined $\tilde{E}\equiv E+j\: \Omega$, representing the fermion energy observed from the inertial frame. In writing Eq.~(\ref{eigenbasis}), the index $\lambda$ represents the set of quantum numbers $\lambda=\{l,k_z,k_\perp,E\}$. Therefore, the solution of Eq.~(\ref{psi}) is
\begin{equation}
\begin{aligned}
        \Psi^\lambda(x)=&\begin{pmatrix}
\left[E+j\Omega+m-k_{z}+ik_{\perp} \right]J_{l}(k_{\perp}\rho) \\
\left[E+j\Omega+m+k_{z}-ik_{\perp} \right]J_{l+1}(k_{\perp}\rho)e^{i\varphi} \\
\left[-E-j\Omega+m-k_{z}+ik_{\perp} \right]J_{l}(k_{\perp}\rho)\\
\left[-E-j\Omega+m+k_{z}-ik_{\perp} \right]J_{l+1}(k_{\perp}\rho)e^{i\varphi}
\end{pmatrix}\\
&\times e^{-(E+j\Omega)t+ik_{z}z+il\varphi}.
\end{aligned}
\end{equation}

Before introducing the explicit form of the fermion propagator in the rotating
environment, it is useful to clarify the assumptions underlying the solution.
Causality requires that $\Omega R<1$, where $R$ denotes the characteristic
transverse size of the rotating system. Typical values extracted from
hydrodynamic and transport simulations for semicentral collisions in the range
$\sqrt{s_{NN}}=2$--20 GeV correspond to
$\Omega \sim 0.02$--$0.10~\mathrm{fm}^{-1}$ and
$R \sim 5$--$8~\mathrm{fm}$, leading to
$\Omega R \lesssim 0.7$, which satisfies the causality constraint. Additionally, we adopt the approximation that the fermion is effectively dragged
by the collective vortical motion during the early stages of the collision.
Within this regime, the azimuthal coordinate follows the rotation according to
$\varphi+\Omega t=0$. This relation should be interpreted as a controlled
approximation valid while the global vortical structure remains coherent and
before transverse expansion significantly alters the velocity profile~\cite{Deng:2016gyh}.

\subsection{Fermion propagator in a rotating environment}

Now, we calculate the propagator of a fermion immersed in a rotating environment. We follow the method developed in Refs.~\cite{Iablokov:2019rpd,Iablokov:2020ypp,Iablokov:2021hgl}. First, consider a Green's function \(G(x,x')\) that satisfies the operator equation
\begin{equation}
\mathcal{\hat{H}}(\partial_x, x) G(x,x') = \delta^4(x-x'),
\label{eq:green_eq}
\end{equation}
where \(\mathcal{\hat{H}}\) is a given Hamiltonian. The Fock-Schwinger method allows us to represent \(G\), the inverse of \(\mathcal{\hat{H}}\), as a proper-time integral
\begin{equation}
G(x,x') = (-i) \int_{-\infty}^{0} d\tau \, U(x,x';\tau),
\label{eq:proper_time_rep}
\end{equation}
where \(U(x,x';\tau)\) is the proper-time evolution operator. The operator \(U(x,x';\tau)\) is defined as the solution of the Schrödinger-like equation
\begin{equation}
i \partial_\tau U(x,x';\tau) = \mathcal{\hat{H}}(\partial_x, x) U(x,x';\tau),
\label{eq:schrodinger}
\end{equation}
satisfying the boundary conditions
\begin{equation}
U(x,x';\infty) = 0, \qquad U(x,x';0) = \delta^4(x-x').
\label{eq:boundary_conditions}
\end{equation}
These conditions ensure that the integral in Eq.~(\ref{eq:proper_time_rep}) produces a causal Green's function.

The formal solution of Eq.~(\ref{eq:schrodinger}) with the given boundary conditions, is given by
\begin{equation}
U(x,x';\tau) = \exp\left[-i\tau \mathcal{\hat{H}}(\partial_x, x)\right] \delta^4(x-x'),
\label{eq:U_operator}
\end{equation}
where the exponential operator acts on the delta function. Substituting this into Eq.~(\ref{eq:proper_time_rep}) yields:
\begin{equation}
G(x,x') = (-i) \int_{-\infty}^{0} d\tau \, \exp\left[-i\tau \mathcal{\hat{H}}(\partial_x, x)\right] \delta^4(x-x').
\label{eq:green_proper_time}
\end{equation}

The \(i\epsilon\) prescription is implicitly understood, that is, \(\mathcal{\hat{H}} \to \mathcal{\hat{H}} - i\epsilon\) to ensure convergence at \(\tau \to -\infty\). This corresponds to the Feynman boundary conditions for the propagator. For a Dirac fermion in a rotating frame, the propagator \(S(x,x')\) is related to the Green's function \(G(x,x')\) of the Klein-Gordon-type operator via
\begin{equation}
S(x,x') =\left[i\gamma^0\left(\partial_t + \Omega \hat{J}_z\right) + i\vec{\gamma}\cdot\vec{\nabla} + m\right] G(x,x').
\label{eq:fermion_prop}
\end{equation}
The key step is to replace $\delta^4(x-x')$ in Eq.~(\ref{eq:green_proper_time}) by the appropriate closure relation satisfied by the eigenfunctions of $\mathcal{\hat{H}}$. To show that the functions $\phi^\lambda(x)$ in Eq.~(\ref{eigenbasis}) satisfy the closure relation, we need to compute
\begin{equation}
    \sum_i\sum_{\ell=-\infty}^\infty\int \frac{dE\,dk_z\,k_\perp dk_\perp}{(2\pi)^3}\,\phi_i^\lambda(x){\phi_i^{\lambda}}^\dagger(x'),
    \label{eq:closure}
\end{equation}
where $\phi_i^\lambda(x)$ is an element of the basis of solutions of $\mathcal{\hat{H}}$ and we have written the solution as 
\begin{eqnarray}
\phi^\lambda(x)=\sum_{i=1}^4\phi_i^\lambda(x), 
\end{eqnarray}
where we take the spinor basis as
\begin{eqnarray}
\phi_1^\lambda(x)&=&\begin{pmatrix}
J_{l}(k_{\perp}\rho) \\
0 \\
0\\
0
\end{pmatrix}
e^{-Et+ik_{z}z+il\varphi},\nonumber\\   \phi_2^\lambda(x)&=&\begin{pmatrix}
0 \\
J_{l+1}(k_{\perp}\rho)e^{i\varphi} \\
0\\
0
\end{pmatrix}
e^{-Et+ik_{z}z+il\varphi},\nonumber\\
\phi_3^\lambda(x)&=&\begin{pmatrix}
0 \\
0 \\
J_{l}(k_{\perp}\rho)\\
0
\end{pmatrix}
e^{-Et+ik_{z}z+il\varphi},\nonumber\\
\phi_4^\lambda(x)&=&\begin{pmatrix}
0 \\
0 \\
0\\
J_{l+1}(k_{\perp}\rho)e^{i\varphi}
\end{pmatrix}
e^{-Et+ik_{z}z+il\varphi}.\nonumber\\
\label{eq:SolPlain}
\end{eqnarray}
Therefore,
\begin{align}
&\sum_{i=1}^4 \sum_{\ell=-\infty}^\infty \int \frac{dE\,dk_z\,dk_\perp k_\perp}{(2\pi)^3} \phi_i^\lambda(x) {\phi_i^\lambda}^\dagger(x') = \nonumber\\
&\quad \sum_{\ell=-\infty}^\infty \int \frac{dE\,dk_z\,dk_\perp k_\perp}{(2\pi)^3} \Big[ J_\ell(k_\perp\rho)J_\ell(k_\perp\rho') e^{i\ell(\varphi-\varphi')} \mathcal{O}^+ \nonumber\\
&\quad + J_{\ell+1}(k_\perp\rho)J_{\ell+1}(k_\perp\rho') e^{i(\ell+1)(\varphi-\varphi')} \mathcal{O}^- \Big]\nonumber\\
&\quad e^{-iE(t-t')} e^{ik_z(z-z')},
\label{eq:closure_rotating}
\end{align}
where the spin projection operators are defined as
\begin{equation}
\mathcal{O}^\pm = \frac{1}{2} \left(1 \pm i \gamma^1 \gamma^2\right).
\label{eq:projectors}
\end{equation}
These operators project onto the two spin states in the plane perpendicular to the rotation axis. Using the orthogonality of the Bessel functions,
\begin{equation}
\int_0^\infty k_\perp dk_\perp \, J_\ell(k_\perp\rho) J_\ell(k_\perp\rho') = \frac{\delta(\rho-\rho')}{\rho},
\label{eq:bessel_orthogonality}
\end{equation}
and the completeness of the Fourier series and Fourier integrals, we obtain:
\begin{equation}
\sum_{i=1}^4 \sum_{\ell=-\infty}^\infty \int \frac{dE\,dk_z\,dk_\perp k_\perp}{(2\pi)^3} \phi_i^\lambda(x) {\phi_i^\lambda}^\dagger(x') = \mathbb{I}_4 \, \delta^4(x-x').
\label{eq:closure_proved}
\end{equation}
This confirms that the set \(\{\phi_i^\lambda(x)\}\) forms a complete orthonormal basis. Returning to the computation of the fermion propagator, from Eqs.~(\ref{eq:green_proper_time}) and~(\ref{eq:fermion_prop}), we have
\begin{eqnarray}
    S(x,x') &=& (-i) \int_{-\infty}^0 d\tau \, \left[i\gamma^0\left(\partial_t + \Omega \partial_\phi\right) + i\vec{\gamma}\cdot\vec{\nabla} + m\right]\nonumber\\
    &\times&\exp\left[-i\tau \mathcal{\hat{H}}\right] \delta^4(x-x'),
\label{eq:S_proper_time}
\end{eqnarray}
where \(\mathcal{\hat{H}}\) is the Klein-Gordon operator in the rotating frame
\begin{equation}
\mathcal{\hat{H}} = -\left(\partial_t + \Omega \partial_\phi\right)^2 + \nabla^2 - m^2.
\label{eq:KG_rotating}
\end{equation}
Using the spectral representation of the delta function in terms of the eigenfunctions \(\phi_i^\lambda(x)\), we write
\begin{equation}
\delta^4(x-x') = \sum_{i=1}^4 \sum_{\ell=-\infty}^\infty \int \frac{dE\,dk_z\,dk_\perp k_\perp}{(2\pi)^3} \phi_i^\lambda(x) {\phi_i^\lambda}^\dagger(x').
\label{eq:delta_spectral}
\end{equation}
Therefore,
\begin{eqnarray}
\exp\left[-i\tau \mathcal{\hat{H}}\right] \delta^4(x-x') &=& \sum_{i,\ell} \int \frac{dE\,dk_z\,dk_\perp k_\perp}{(2\pi)^3}\nonumber\\
&\times&e^{-i\tau \mathcal{H}} \phi_i^\lambda(x) {\phi_i^\lambda}^\dagger(x'),
\label{eq:H0_action}
\end{eqnarray}
where on the right-hand side of Eq.~(\ref{eq:H0_action}), $\mathcal{H} = \tilde{E}^2 - k_z^2 - k_\perp^2 - m^2$ is now an eigenvalue. We write Eq.~(\ref{eq:S_proper_time}) introducing the matrix
\begin{widetext}
    \begin{eqnarray}
       &&\mathcal{D} \equiv i\gamma^0\left(\partial_t+ \Omega\hat{J}_z \right) +i\Vec{\gamma}\cdot\Vec{\nabla}+m=\begin{pmatrix}
        i\left(\partial_t+\Omega\hat{j}\right)+m & 0 &-i\partial_z & -\mathcal{P}_-\\
         0 & i\left(\partial_t+\Omega\hat{j}\right)+m &-\mathcal{P}_+ & i\partial_z\\
         i\partial_z& \mathcal{P}_- & -i\left(\partial_t+\Omega\hat{j}\right)+m&0\\
         \mathcal{P}_+&-i\partial_z&0&-i\left(\partial_t+\Omega\hat{j}\right)+m
    \end{pmatrix},
\end{eqnarray}
where $\mathcal{P}_\pm=k_x\pm ik_y$ satisfies \begin{eqnarray}
    \mathcal{P}_\pm J_{\ell} (k_\perp \rho)e^{i \ell\varphi}e^{-iEt}e^{ik_zz}=\pm i k_\perp J_{\ell\pm1}(k_\perp \rho)e^{i (\ell\pm1)\varphi}e^{-iEt}e^{ik_zz},
\end{eqnarray}
and
\begin{eqnarray}
    ( i\left(\partial_t+\Omega\hat{j}\right)+m) J_{\ell} (k_\perp \rho)e^{i \ell\varphi}e^{-iEt}e^{ik_zz}=(\Tilde{E}+m) J_{\ell} (k_\perp \rho)e^{i \ell\varphi}e^{-iEt}e^{ik_zz},
\end{eqnarray}
\begin{eqnarray}
   i\partial_z J_{\ell} (k_\perp \rho)e^{i \ell\varphi}e^{-iEt}e^{ik_zz}=k_z J_{\ell} (k_\perp \rho)e^{i \ell\varphi}e^{-iEt}e^{ik_zz},
\end{eqnarray}

With all these elements and after integrating over the proper time $\tau$ in Eq.~(\ref{eq:S_proper_time}), the fermion propagator can be expressed as
\begin{equation}
   S(x,x') =(-i)\sum_{l=-\infty}^{\infty}\int \frac{dE\;dk_{z}\;dk_{\perp} k_{\perp}}{(2\pi)^{3}} \Phi(x,x') \frac{1}{\tilde{E}^{2}-k_{z}^{2}-k_{\perp}^{2}-m^{2}+i\epsilon},
   \label{Prop1}
\end{equation}
where $\Phi(x,x')$ contains the action of the Dirac operator on the eigenfunctions. To see the explicit form of $\Phi(x,x')$, we first notice that it can be written in terms of the spin projection operators $\mathcal{O}^{\pm}$ as
\begin{equation}
    \Phi(x,x') = \mathcal{D}^{+}J_{l}(k_{\perp}\rho)J_{l}(k_{\perp}\rho')e^{il(\varphi-\varphi')}e^{-iE(t-t')}e^{ik_{z}(z-z')} + \mathcal{D}^{-}J_{l+1}(k_{\perp}\rho)J_{l+1}(k_{\perp}\rho')e^{i(l+1)(\varphi-\varphi')}e^{-iE(t-t')}e^{ik_{z}(z-z')},
    \label{capitalphi}
\end{equation}
with the operators $\mathcal{D}^{\pm}\equiv\mathcal{D}\mathcal{O}^{\pm}$ explicitly given by
\begin{equation}
\mathcal{D}^{+}=\begin{pmatrix}
        i\left(\partial_{t}+\Omega\hat{J}_{z}\right)+m & 0 &-i\partial_{z} & 0\\
         0 & 0 &-(i\partial_{x}-\partial_{y}) & 0\\
         i\partial_{z}& 0 & -i\left(\partial_{t}+\Omega\hat{J}_{z}\right)+m&0\\
         (i\partial_{x}-\partial_{y})&0&0&0
    \end{pmatrix},
\end{equation}
\begin{equation}
\mathcal{D}^{-}=\begin{pmatrix}
        0 & 0 &0 & -(i\partial_{x}+\partial_{y})\\
         0 & i\left(\partial_{t}+\Omega\hat{J}_{z}\right)+m &0 & i\partial_{z}\\
        0& (i\partial_{x}+\partial_{y}) &0&0\\
         0&-i\partial_{z}&0&-i\left(\partial_{t}+\Omega\hat{J}_{z}\right)+m
    \end{pmatrix}.
\end{equation}
Applying these operators to the Bessel functions we obtain
\begin{equation}
\begin{aligned}
&\mathcal{D}^{+}J_{l}(k_{\perp}\rho)J_{l}(k_{\perp}\rho')e^{il(\varphi-\varphi')}e^{-iEt}e^{ik_{z}z}=
&\begin{pmatrix}
       (\tilde{E}+m)J_{l}J_{l}'e^{il(\varphi-\varphi')} & 0 &-k_{z}J_{l}J_{l}'e^{il(\varphi-\varphi')} & 0\\
         0 & 0 &-ik_{\perp}J_{l+1}J_{l}'e^{il(\varphi-\varphi')+i\varphi} & 0\\
         k_{z}J_{l}J_{l}'e^{il(\varphi-\varphi')}& 0 & (-\tilde{E}+m)J_{l}J_{l}'e^{il(\varphi-\varphi')}&0\\
         ik_{\perp}J_{l-1}J_{l}'e^{il(\varphi-\varphi')-i\varphi}&0&0&0
    \end{pmatrix}e^{-iEt}e^{ik_{z}z},
\end{aligned}
\end{equation}
and
\begin{equation}
\begin{aligned}
\mathcal{D}^{-}J_{l+1}(k_{\perp}\rho)J_{l+1}(k_{\perp}\rho')e^{i(l+1)(\varphi-\varphi')}e^{-iEt}e^{ik_{z}z}=
&\begin{pmatrix}
        0 & 0 &0 & ik_{\perp}J_{l}J_{l+1}'e^{il(\varphi-\varphi')-i\varphi'}\\
         0 & (\tilde{E}+m)J_{l+1}J_{l+1}'e^{i(l+1)(\varphi-\varphi')} &0 & k_{z}J_{l+1}J_{l+1}'e^{i(l+1)(\varphi-\varphi')}\\
        0& -ik_{\perp}J_{l}J_{l+1}'e^{il(\varphi-\varphi')-i\varphi'} &0&0\\
         0&-k_{z}J_{l+1}J_{l+1}'e^{i(l+1)(\varphi-\varphi')}&0&(-\tilde{E}+m)J_{l+1}J_{l+1}'e^{i(l+1)(\varphi-\varphi')}
    \end{pmatrix}\\
    &\times e^{-iEt}e^{ik_{z}z},
\end{aligned}
\end{equation}
where we have used the notation $J_{l}\equiv J_{l}(k_{\perp}\rho)$ and $J_{l}'\equiv J_{l}(k_{\perp}\rho')$. To perform the sum over $l$, we employ the Anger-Jacobi identity
\begin{equation}
    \sum_{l=-\infty}^{\infty} J_{l}(x)e^{ily} = e^{ix\sin y}.
\end{equation}
After applying the rigid rotation approximation $\varphi-\varphi'+\Omega t =0$ and making the change of variables $\rho' = R - r/2$, $\rho = R + r/2$, we obtain for the first term in Eq.~(\ref{capitalphi})
\begin{equation}
\begin{aligned}
\sum_{l=-\infty}^{\infty}&\mathcal{D}^{+}J_{l}(k_{\perp}\rho)J_{l}(k_{\perp}\rho')e^{il(\varphi-\varphi')}e^{-iEt}e^{ik_{z}z} = e^{-i(E-\Omega/2)(t-t')}e^{ik_{z}(z-z')}\\
&\times\Big{[}(\gamma^{0}E-\gamma^{3}k_{z}+m)J_{0}(k_{\perp}r)-i(\gamma^{1}\cos\varphi+\gamma^{2}\sin\varphi)k_{\perp}J_{1}(k_{\perp}r)\Big{]}\mathcal{O}^{+}.
\end{aligned}
\end{equation}
Similarly, for the second term in Eq.~(\ref{capitalphi}) we find
\begin{equation}
\begin{aligned}
\sum_{l=-\infty}^{\infty}&\mathcal{D}^{-}J_{l+1}(k_{\perp}\rho)J_{l+1}(k_{\perp}\rho')e^{i(l+1)(\varphi-\varphi')}e^{-iEt}e^{ik_{z}z} = e^{-i(E+\Omega/2)(t-t')}e^{ik_{z}(z-z')}\\
&\times\Big{[}(\gamma^{0}E-\gamma^{3}k_{z}+m)J_{0}(k_{\perp}r)-i(\gamma^{1}\cos\varphi+\gamma^{2}\sin\varphi)k_{\perp}J_{1}(k_{\perp}r)\Big{]}\mathcal{O}^{-}.
\end{aligned}
\end{equation}
Notice that with the rigid rotation approximation, the propagator is translationally invariant and can be simply Fourier transformed. Therefore, substituting these expressions back into Eq.~(\ref{Prop1}), the propagator becomes
\begin{equation}
\begin{aligned}
S(x,x') =&(-i)\int \frac{dE\;dk_{z}\;dk_{\perp} k_{\perp}}{(2\pi)^{3}} \frac{e^{-i(E-\Omega/2)(t-t')}e^{ik_{z}(z-z')}}{E^{2}-k_{z}^{2}-k_{\perp}^{2}-m^{2}+i\epsilon}\Big{[}(\gamma^{0}E-\gamma^{3}k_{z}+m)J_{0}(k_{\perp}r)-i\gamma_{\perp}\cdot k_{\perp}J_{1}(k_{\perp}r)\Big{]}\mathcal{O}^{+}\\
&+(-i)\int \frac{dE\;dk_{z}\;dk_{\perp} k_{\perp}}{(2\pi)^{3}} \frac{e^{-i(E+\Omega/2)(t-t')}e^{ik_{z}(z-z')}}{E^{2}-k_{z}^{2}-k_{\perp}^{2}-m^{2}+i\epsilon}\Big{[}(\gamma^{0}E-\gamma^{3}k_{z}+m)J_{0}(k_{\perp}r)-i\gamma_{\perp}\cdot k_{\perp}J_{1}(k_{\perp}r)\Big{]}\mathcal{O}^{-},
\end{aligned}
\end{equation}
where we have defined $\gamma_{\perp}\cdot k_{\perp} \equiv (\gamma^{1}\cos\varphi+\gamma^{2}\sin\varphi)k_{\perp}$.
\end{widetext}
Performing the Fourier transform to momentum space, after integrating over the angular coordinates and using the integral representations for the Bessel functions, we arrive at the compact expression
\begin{equation}
\begin{aligned}
S(p)=&\frac{\gamma^{0}(p_{0}+\Omega/2)-\gamma^{3}p_{3}-\gamma_{\perp}\cdot p_{\perp} +m}{(p_{0}+\Omega/2)^{2}-p_{3}^{2}-p_{\perp}^{2}-m^{2}+i\epsilon}\mathcal{O}^{+}\\
&+\frac{\gamma^{0}(p_{0}-\Omega/2)-\gamma^{3}p_{3}-\gamma_{\perp}\cdot p_{\perp} +m}{(p_{0}-\Omega/2)^{2}-p_{3}^{2}-p_{\perp}^{2}-m^{2}+i\epsilon}\mathcal{O}^{-}.
\label{PropFinal}
\end{aligned}
\end{equation}

Notice that when $\Omega\to 0$ we recover the usual vacuum fermion propagator. Since the present derivation is performed in vacuum, to use this propagator in a finite temperature and baryon chemical potential environment in equilibrium, we should perform the replacement $p_{0}\to i\tilde{\omega}_{n}+\mu_B$ where $\tilde{\omega}_{n}=(2n+1)\pi T$ are Matsubara frequencies for fermions. Equation~(\ref{PropFinal}) represents our approximation for the fermion propagator in a rigidly rotating environment with cylindrical geometry. We now proceed to use this propagator to compute the relaxation time for the fermion spin to align with the angular velocity in the rotating corona medium.

\section{Interaction rate for the \texorpdfstring{$\Lambda$}{Lambda} spin to align with the angular velocity within a rotating hadron cloud}\label{secIV}

In a hadron medium in thermal equilibrium at temperature $T$ and baryon chemical potential $\mu_B$, the interaction rates $\Gamma^\pm$ for $\Lambda$ hyperons with spin projections $s=\pm 1/2$ along $\vec{\Omega}$ and four-momentum $P=(p_0,\Vec{p})$ can be written as
\begin{eqnarray}
    \Gamma^\pm(p_0)= \tilde{f}(p_0 - \mu_B \mp \Omega / 2)\text{Tr}\left[\text{Im}\Sigma^\pm\right],
\label{gammadef}
\end{eqnarray}
where $\Sigma^\pm$ is the self-energy of an aligned (+) or anti-aligned (-) $\Lambda$. In the RMF model, the Feynman diagram one-loop self-energy of $\Lambda$ is depicted in Fig.~\ref{loop}, whose explicit expression is
\begin{eqnarray}
\Sigma^\pm(P)=g_\Lambda^2 T\sum_n\int\frac{d^3k}{(2\pi)^3} S^\pm(P-K) \Delta^*(K),    
\end{eqnarray}
where $S^\pm$ are the $\Lambda$ spin up and down components of the propagator in a rotating environment and $\Delta^*$ is the effective $\sigma$ propagator in the thermal medium. The four-momenta are $P=(i\Tilde{\omega},\vec{p})$ for the fermion and $K=(i\omega_n,\vec{k})$ for the $\sigma$ with $\omega_n$ being the $\sigma$ Matsubara frequencies. Also, $\Delta^*$ is given by~\cite{Ayala:2025bzk}
\begin{eqnarray}
\Delta^*(p_0,p)=\frac{-1}{P^2-M_\sigma^2-p_0F(x)-i\pi A(x)\theta(p^2-p_0^2)},
\label{Propagator}
\end{eqnarray}
where 
\begin{eqnarray}
    P^2=p_0^2-p^2,\ \ \ \ M_\sigma^2=m_\sigma^2+M_T^2,
   \end{eqnarray}

\begin{eqnarray}
\!\!\!\!\!\!F(x)&\!\!\!=\!\!\!&\Bigg{[} 3 \left(2 x^2+1\right) x \lambda_1- \lambda_2\nonumber\\
&\!\!\!+\!\!\!& \left(5 x^2+1\right)\gamma_T\Bigg{]}x\ln \left(\frac{x+1}{x-1}\right)\nonumber\\
    &\!\!\!-\!\!\!& \Bigg{[}2  \frac{  \left(5 x^2+1\right)}{\sqrt{3 x^2+1}}\gamma_T+\frac{x \left(10 x^2+7\right)}{\sqrt{3 x^2+1}} \lambda_1\Bigg{]}\nonumber\\
    &\!\!\!\times\!\!\!&x^2\ln \left(\frac{x^2+\sqrt{3 x^2+1}+1}{x^2-\sqrt{3 x^2+1}+1}\right)\nonumber\\
   &\!\!\!+\!\!\!&\left[4 x \left(2 x^2+1\right) \lambda_1+2\lambda_2+\frac{2}{3} \left(15 x^2-2\right)\gamma_T\right],
    \label{sigmaselfF}
\end{eqnarray}

    and
  
   \begin{eqnarray}
        A(x)
        &=&\gamma_T \left(5 x^2+1\right) \left(x-\frac{2 x^2  }{ \sqrt{3 x^2+1}}\right)\nonumber\\
        &+&\lambda_1x^2 \left(3(2x^2+1) -\frac{x(10x^2+7)}{\sqrt{3x^2 +1}}\right)\nonumber\\
        &-&\lambda_2,
    \end{eqnarray}
and
\begin{eqnarray}
    M_T^2&\equiv&\frac{2 g_\sigma^2 M_N T}{\pi^2} K_1\left(\frac{M_N}{T}\right) \cosh \left(\frac{\mu_B }{T}\right),\nonumber\\
    \gamma_T&\equiv&  \frac{2 g_\sigma^2M_N}{\pi^2}  K_1\left(\frac{M_N}{T}\right) \sinh \left(\frac{\mu_B }{T}\right),\nonumber\\
  \lambda_1&\equiv&\frac{2 g_\sigma^2T}{\pi^2} e^{-\frac{M_N}{T}} \cosh \left(\frac{\mu_B }{T}\right),\nonumber\\
    \lambda_2&\equiv&\frac{2 g_\sigma^2T}{\pi^2} e^{-\frac{M_N}{T}} \sinh \left(\frac{\mu_B }{T}\right),
\end{eqnarray}
where $x\equiv p_0/p$. In order to compute the sum over the Matsubara frequencies, it is convenient to  express it in terms of an integral involving products of the propagator spectral densities for
the $\sigma$ and the $\Lambda$ in a rotating environment, $\tilde{\Delta}$, with the replacement  $p_0\to i\tilde{\omega}_n+\mu_B$~\cite{LeBellac}. The latter is naturally decomposed into the two spin-projection components Let us define 
\begin{equation}
S_m(i\omega)=T\sum_n\Delta^*(i\omega_n)\tilde{\Delta}(i(\omega-\tilde{\omega}_n)).   \end{equation}
Now, we can write the imaginary part of $S_m$ by introducing $\rho$ and $\rho_F$, the spectral densities for the $\sigma$-boson and the fermion, respectively, as follows
\begin{eqnarray}
    \text{Im}(S_m)\!\!\!&=&\!\!\!\pi(e^{\beta(p_0-\mu_B - \Omega / 2)}+1)\int_{-\infty}^{\infty}\int_{-\infty}^{\infty}\frac{dk_0}{2\pi}\frac{dp'_0}{2\pi}f(k_0)\nonumber\\
    \!\!\!\!\!\!\!&\times&\!\!\tilde{f}(p'_0-\mu_B \mp \Omega / 2) \delta(p_0-k_0-p'_0)\nonumber \\
    & \times &\rho(k_0,k)\rho_F(p'_0,p-k),
\label{Matsu1}
\end{eqnarray}
where $f(k_0)$ is the Bose-Einstein distribution. The spectral density $\rho$ is obtained from the imaginary part of $\Delta^*(i\omega_n)$ after the analytic continuation $i\omega_n\rightarrow k_0+i\epsilon$ and contains the discontinuities of the $\sigma$ propagator across the real $k_0$ axis, as is described in Ref.~\cite{Ayala:2025bzk}. On the other hand, the fermion spectral density is
\begin{equation}
        \rho_F(p_0',p)=-2\pi\delta \left((p_0'\pm\Omega/2)^2-p^2-M_\Lambda^2\right).\\
    \label{rho}
    \end{equation}
The trace in  Eq.~(\ref{gammadef}) can be readily computed with the result
\begin{eqnarray}
\text{Tr}\left[\left( \gamma^0(p_0\pm\Omega/2)-\Vec{\gamma}\cdot \Vec{p} +M_\Lambda \right)\mathcal{O}^\pm \right]&=&-8M_\Lambda.
\end{eqnarray}
The delta functions in Eqs.~\eqref{Matsu1} and~\eqref{rho} impose a constraint on the $\sigma$ energies, restricting them to the spacelike region, $|x|<1$. Therefore, the contribution from the $\sigma$ spectral density is
\begin{eqnarray}
\!\!\!\!\!\!\!\!\!\beta(p_0,p)&\!\!\!=\!\!\!&2\pi p_0 A(x)\theta\left(1-x^2\right)\nonumber\\
&\!\!\!\times\!\!\!&\left[\left(P^2\!-\!M_\sigma^2\!-\! p_0 F\left(x\right)\right)^2\!\!-\Big{(}p_0A\left(x\right)\pi\Big{)}^2\right]^{-1}\!\!\! ,
\end{eqnarray}
and the expression for the interaction rate, Eq.~(\ref{gammadef}), becomes
\begin{IEEEeqnarray}{rCl}
        \Gamma^\pm (p_0) & = & \frac{g_\Lambda^2M_\Lambda\pi}{2} \int \frac{d^3 k}{(2\pi)^3} \int_{-\infty} ^\infty \frac{dk_0}{2\pi} \int_{-\infty} ^\infty dp_0 ^\prime \nonumber\\
       &\times &
       f(k_0)\left(8\rho(k_0)  \right)\nonumber \\
&  \times &\tilde{f}(p_0 ^\prime - \mu_B \mp \Omega / 2) \delta(p_0 - k_0 - p_0 ^\prime)\nonumber\\  
&\times & \delta\left(\left(p_0 ^\prime \pm \Omega/2\right)^2 - E^2\right),
    \label{gama0}
\end{IEEEeqnarray}
with 
\begin{equation}
    E^2=|\vec{p}-\vec{k}|^2-M_\Lambda^2.
\end{equation}
Notice that
\begin{eqnarray}
    &&\delta\left(\left(p_0' \pm \Omega/2\right)^2 - E^2\right)=\nonumber\\
    &&\frac{1}{2E}\Big[\delta(p_0'\pm\Omega/2 - E)+\delta(p_0^\prime\pm\Omega/2 + E) \Big].
\end{eqnarray}
Therefore, we can integrate Eq.~(\ref{gama0}) over $p_0^\prime$ to obtain
\begin{IEEEeqnarray}{rCl}
        \Gamma^\pm (p_0) & = & \frac{g_\Lambda^2M_\Lambda\pi}{2} \int \frac{d^3 k}{(2\pi)^3} \int_{-\infty} ^\infty \frac{1}{2E}\frac{dk_0}{2\pi} f(k_0)8\rho_{T}(k_0) \nonumber\\   
&  \times & \Big[\tilde{f}(E - \mu_B\mp\Omega)\delta(p_0 - k_0 - E \pm\Omega/2)\nonumber\\
&+ &\tilde{f}(-E - \mu_B\mp\Omega)\delta(p_0 - k_0 + E \pm\Omega/2)\Big].
    \label{gama1}
\end{IEEEeqnarray}
The first term in Eq.~(\ref{gama1}) corresponds to the rate of production of rotating and thermalized $\Lambda$ hyperons originating from the scattering of originally non-thermalized and non-rotating $\Lambda$s as a result of dispersion with medium $\sigma$s. The second term corresponds to the rate of production of $\overline{\Lambda}$s. Therefore, for our present purposes, we only consider the contribution of the first term and write
\begin{IEEEeqnarray}{rCl}
        \Gamma^\pm (p_0) & = & \frac{g_\Lambda^2M_\Lambda\pi}{2}\!\!\! \int_0^\infty\!\!\frac{k^2dk\;d(\cos\theta)d\phi}{(2\pi)^3}\!\! \int_{-\infty} ^\infty\!\!\frac{1}{2E}\frac{dk_0}{2\pi} 8\rho_{T}(k_0)\nonumber \\
       &\times &   \delta(p_0 - k_0 - E \pm\Omega/2) f(k_0) \tilde{f}(E -\mu_B\mp\Omega).
    \label{gama2}
\end{IEEEeqnarray}

The remaining delta function imposes a kinematical restriction for the  $k_0$ integration, which
translates into integrations only over the regions $\mathcal{R^\pm}$, defined below. After integrating the angle $\theta$ between $\vec{p}$ and $\vec{k}$ and the azimuthal angle $\phi$, and using that $E^2=p^2+\ m^2=|\vec{p}-\vec{k}|^2-m^2=p^2+k^2-2pk\cos{\theta}+m^2$, we obtain
\begin{IEEEeqnarray}{rCl}
\Gamma^\pm (p_0)  & = & \frac{g_\Lambda^2 M_\Lambda\pi}{2}\int_0 ^\infty \frac{dk\,k^2}{(2\pi)^3}\int_{\mathcal{R^\pm}}dk_0 \frac{f(k_0)}{2pk}\nonumber\\
&\times&8\rho_{T}(k_0)  \tilde{f}\left(p_0 - k_0 - \mu_B \mp \Omega/2\right),\IEEEeqnarraynumspace
    \label{GammaT1}
\end{IEEEeqnarray}

where $\mathcal{R^\pm}$ are the regions defined by
\begin{equation}
\begin{aligned}
  k_0 &\leq p_0\pm\Omega/2 - \sqrt{(p-k)^2+M_\Lambda^2},\\
 k_0 &\geq p_0\pm\Omega/2 - \sqrt{(p+k)^2+M_\Lambda^2}.
\end{aligned}
\end{equation}
\begin{figure}[t]
    \centering
    \includegraphics[width=\linewidth]{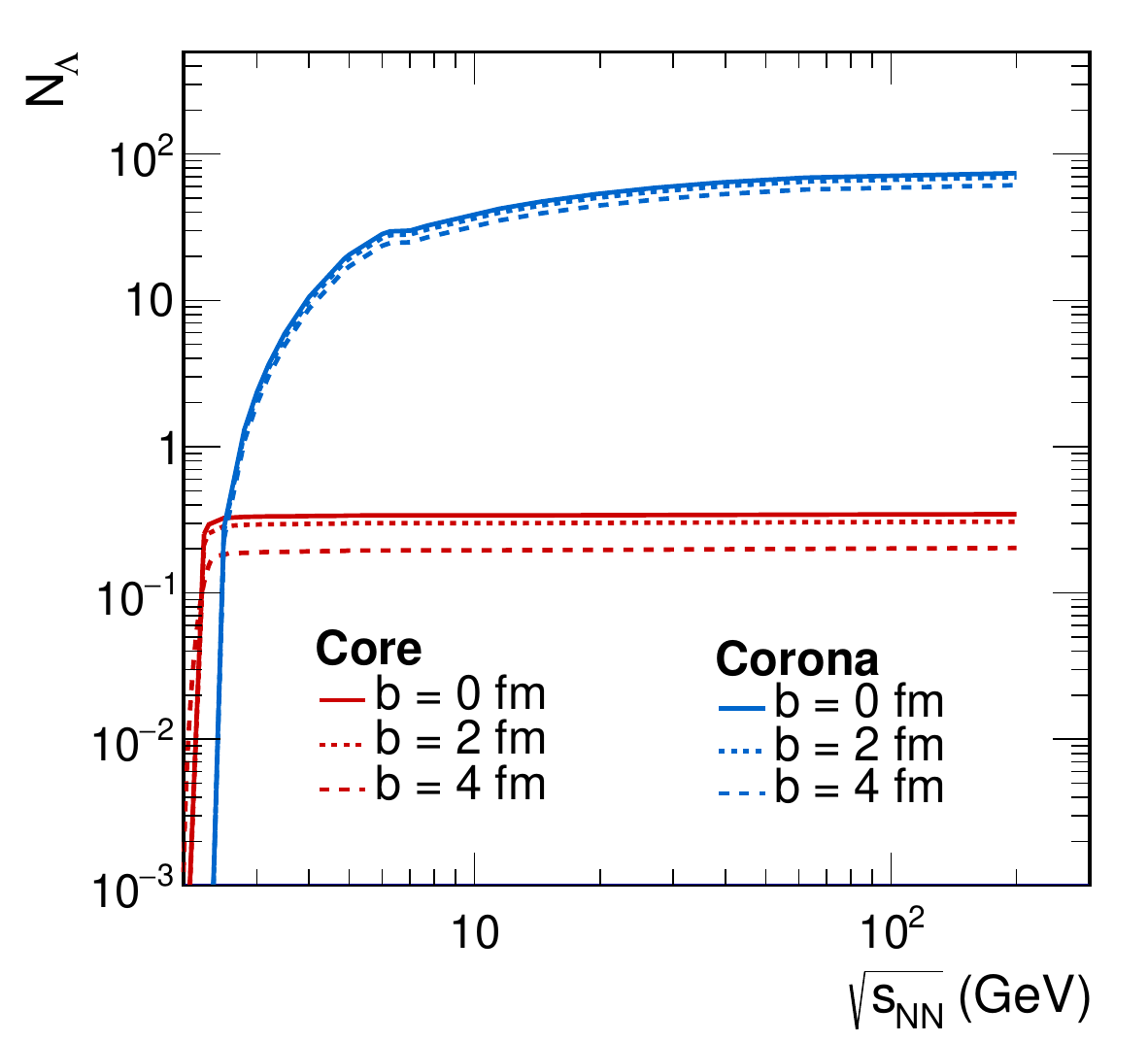}
    \caption{Number of $\Lambda$s produced in the corona (blue) and the core (red) as function of the collision energy for impact parameters $b$=0, 2, 4 fm using $n_c=3.5$ fm$^{-2}$.}
    \label{fig:Nlambda}
\end{figure}
The total rate for aligning the quark spin with the angular velocity is therefore given by the difference between the rates of populating spin projections parallel and antiparallel to the angular velocity. This is obtained by integrating the difference between $\Gamma^+$ and $\Gamma^-$ from Eq.~(\ref{GammaT1}) over the available phase space
\begin{equation}
    \Gamma=V\int\frac{d^3p}{(2\pi)^3}\left[\Gamma^+(p_0)-\Gamma^-(p_0)\right],
    \label{gammaTot2}
\end{equation}
where $V$ is the volume of the collision region.

\section{Computing the number of \texorpdfstring{$\Lambda$}{Lambda}'s produced from the core and the corona}\label{secV}

To evaluate the global $\Lambda$ polarization within the Core--Corona framework, it is necessary to determine not only the intrinsic polarization generated in each subregion but also the relative abundance of $\Lambda$ hyperons produced in the core and in the corona. We estimate these yields using a Glauber description of the participant density in the transverse plane. This construction provides a natural geometric criterion to separate the dense region, where QGP formation is expected, from the dilute region, where hadron production processes dominate.

\begin{figure}[t!]
\centering
\includegraphics[width=\linewidth]{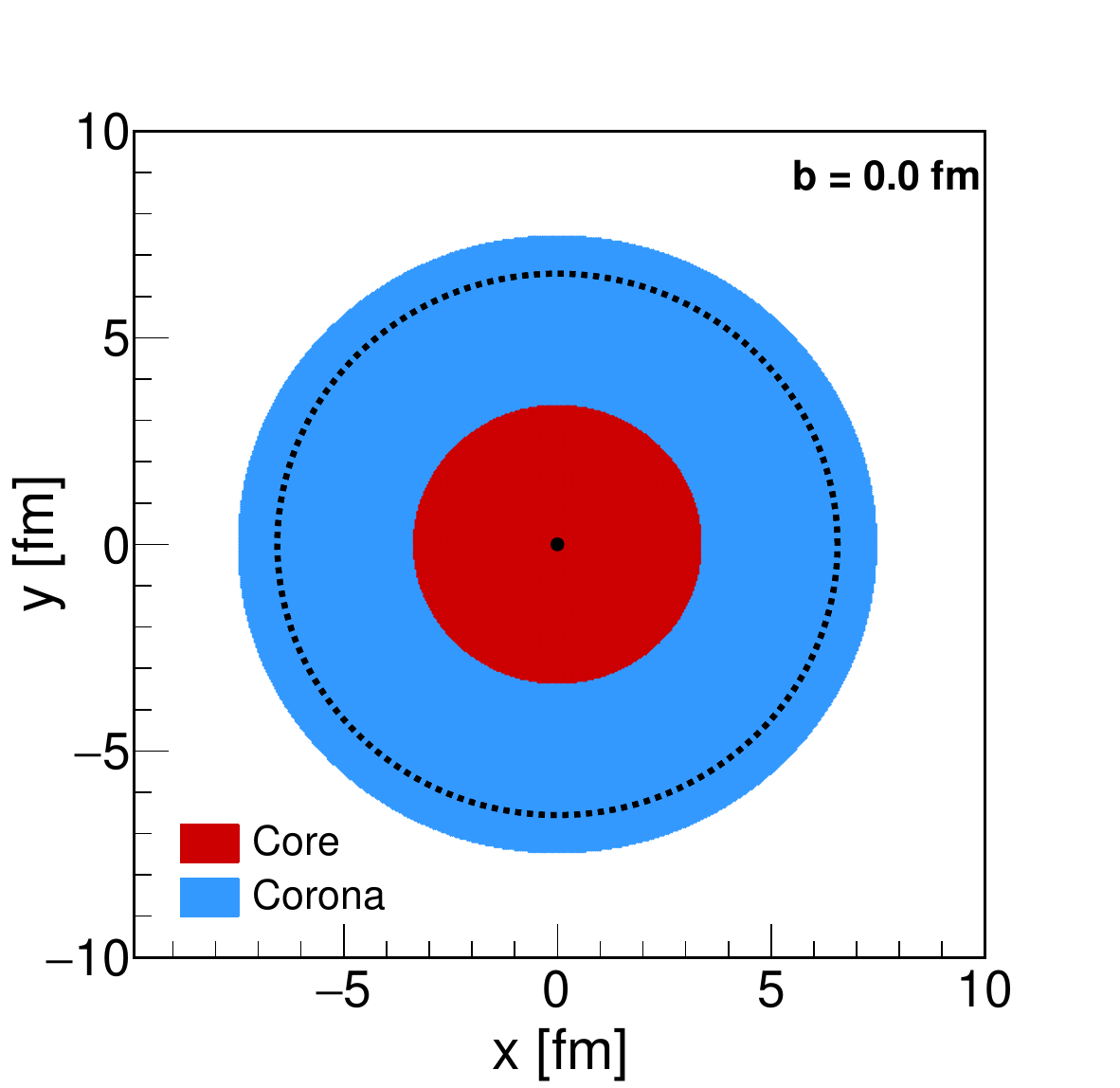}
\includegraphics[width=\linewidth]{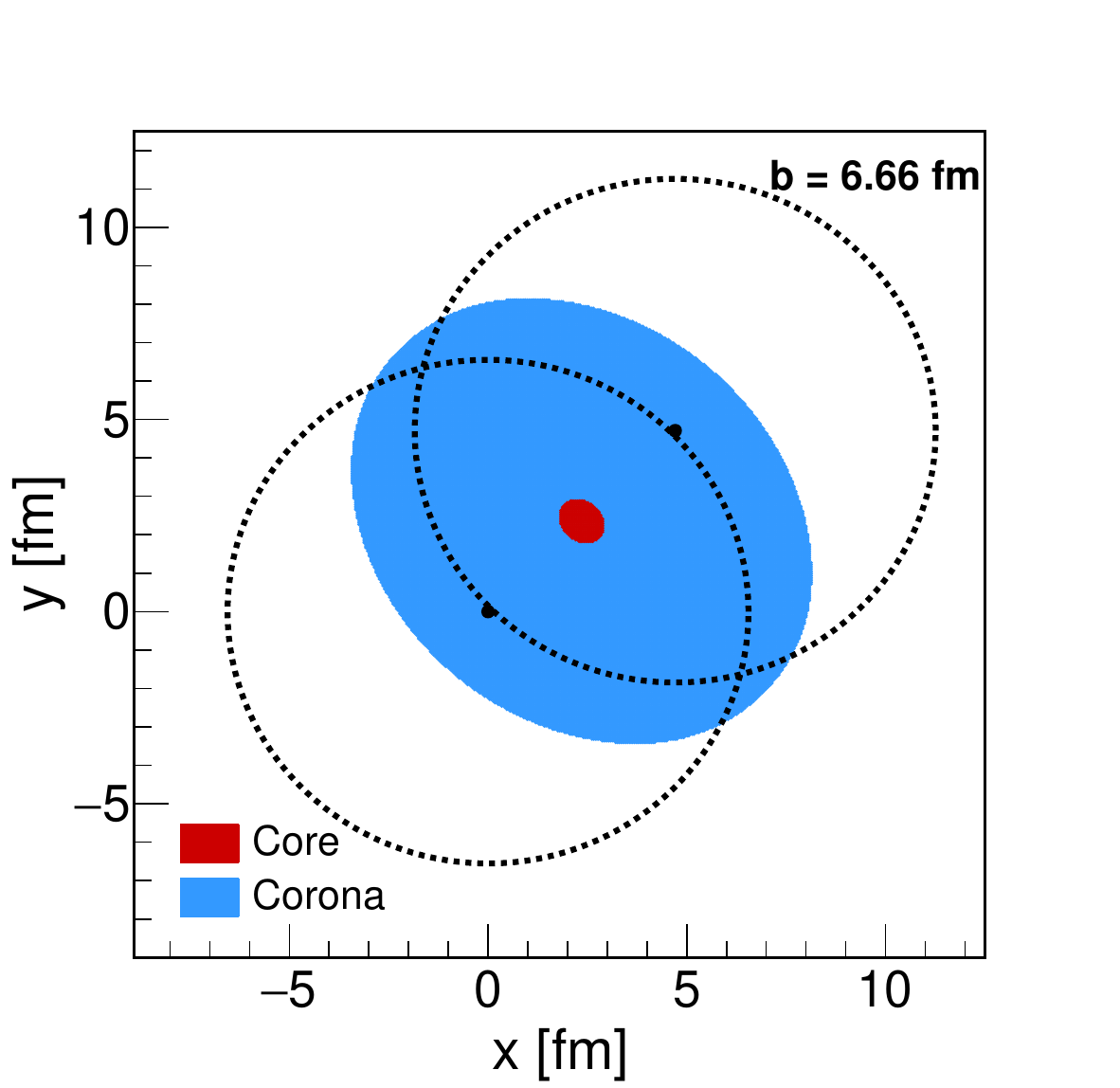}
\caption{Core (red) and corona (blue) regions for $\sqrt{s_{NN}}=$ 200 GeV Au+Au collisions and two impact parameters $b=0$ (top) and $b=6.66$ fm (bottom).
In both panels, the dotted lines represent the contours of the colliding ions and the black dots their corresponding centers. Notice that the core region diminishes as the impact parameter increases.}
\label{fig:coreregion}
\end{figure}

From Eq.~(\ref{numbers}), the remaining ingredient needed to compute the total global $\Lambda$ polarization, including both core and corona contributions, is the number of $\Lambda$ hyperons produced in each subregion. We calculate the number of these particles using a Glauber model, introducing the density of participant nucleons in the collision $n_p(\vec{s},\vec{b})$ at a position $\vec{s}$ in the transverse plane of the collision as a function of the impact parameter vector $\vec{b}$. The density is expressed in terms of the thickness functions $T_A$ and $T_B$ of the colliding system $A+B$
\begin{eqnarray}
    n_p(\Vec{s},\Vec{b})&=&T_A(\Vec{s})\left[1-e^{-\sigma_{NN}(\sqrt{s_{NN}})T_B(\Vec{s}-\Vec{b})} \right]\nonumber\\
    &+&T_B(\Vec{s}-\Vec{b})\left[1-e^{-\sigma_{NN}(\sqrt{s_{NN}})T_A(\Vec{s})} \right],
    \label{GM}
    \end{eqnarray}
where the thickness function is given by
    \begin{eqnarray}
    T_A(\Vec{s}) = \int_{-\infty}^\infty dz \rho_A(z,\Vec{s}) =  \int_{-\infty}^\infty dz \frac{\rho_0}{1+e^{\frac{r-R_A}{a}}},
    \label{thickness}
    \end{eqnarray}
and is taken as a Woods–Saxon distribution with a skin depth
$a$ = 0.523 fm and a radius $R$ = 6.554 fm for the Au nucleus, and $\sigma_{NN}$ is the $N + N$ cross-section, which is collision energy-dependent. We introduce the critical density of participants $n_c$, such that the production of QGP happens when $n_p>n_c$ and take $n_c=3.5 $ fm$^{-2}$. Then the number of $\Lambda$s from the core, is proportional to the number of participant nucleons in the
collision above this critical value and is explicitly given by
\begin{eqnarray}
    N_{p\;QGP}=\int d^2s\;n_p(\Vec{s},\Vec{b})\;\theta\left[ n_p(\Vec{s},\Vec{b})-n_c\right].
\end{eqnarray}
\begin{figure}[t]
    \centering
    \includegraphics[width=\linewidth]{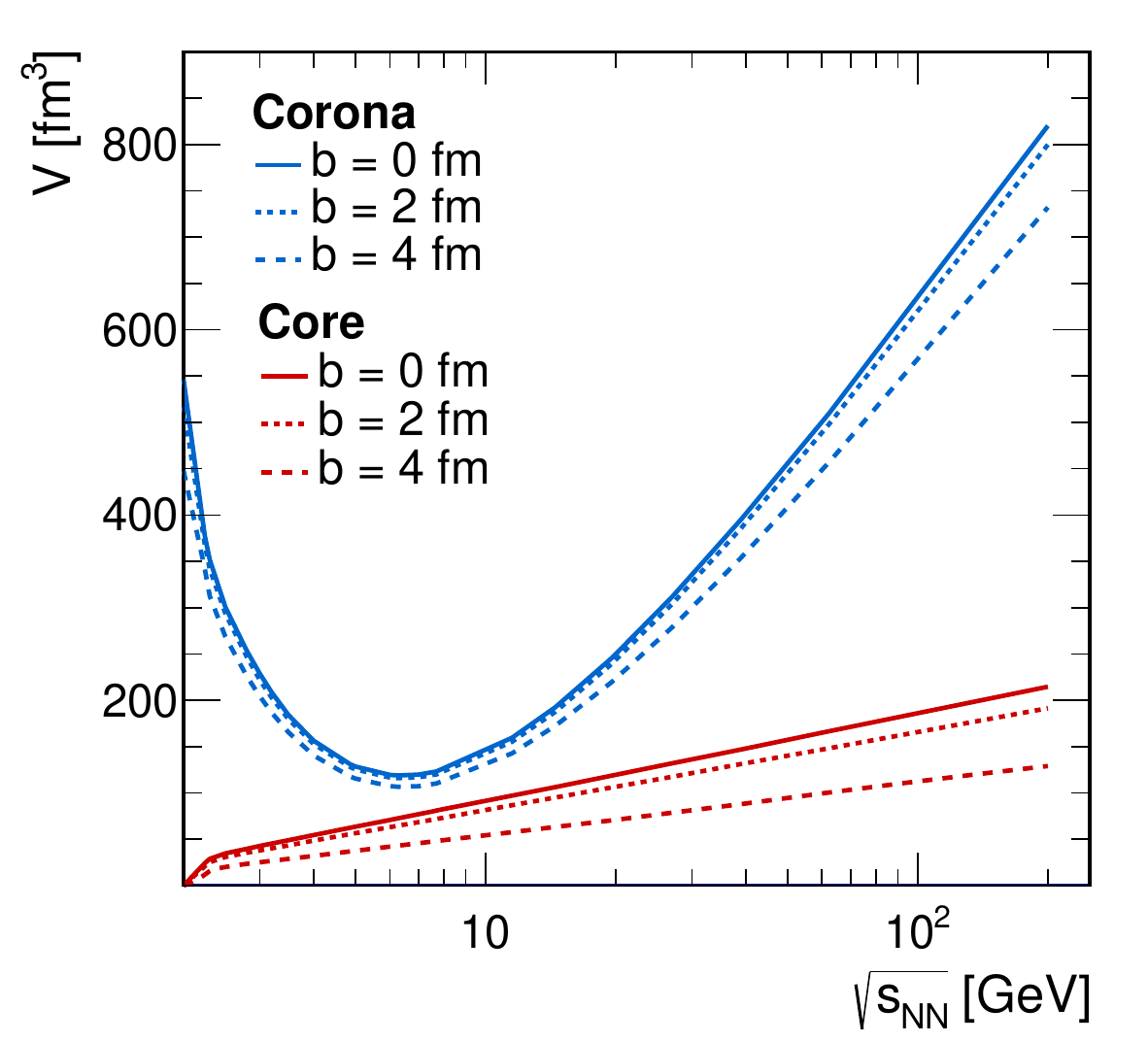}
    \caption{Core (red) and corona (blue) volumes as functions of $\sqrt{s_{NN}}$ for impact parameters b = 0, 2, 4 fm. }
    \label{fig:lambdaVol}
\end{figure}

With this information at hand, we can estimate the average number of strange quarks produced in the QGP, and thus the number of $\Lambda$s, as a quantity
that scales with the number of participants in the collision, as
\begin{equation}
<s>= N_{\Lambda\;QGP} = c N_{p\;QGP}
\end{equation}
where $c$ is in the range $0.001\leq c \leq 0.005$ \cite{Letessier:1996ad}. In this work, we use $c$ = 0.0025. On the other hand, the number of $\Lambda$s produced in the corona can be computed as 
\begin{equation}
    N_{\Lambda\;REC} = \sigma_{NN}^\Lambda\int d^2s\;T_B(\Vec{s}-\Vec{b})T_A(\Vec{s})\;\theta\left[n_c - n_p(\Vec{s},\Vec{b})\right],
\end{equation}
\begin{figure}[t]
    \centering
    \includegraphics[width=\linewidth]{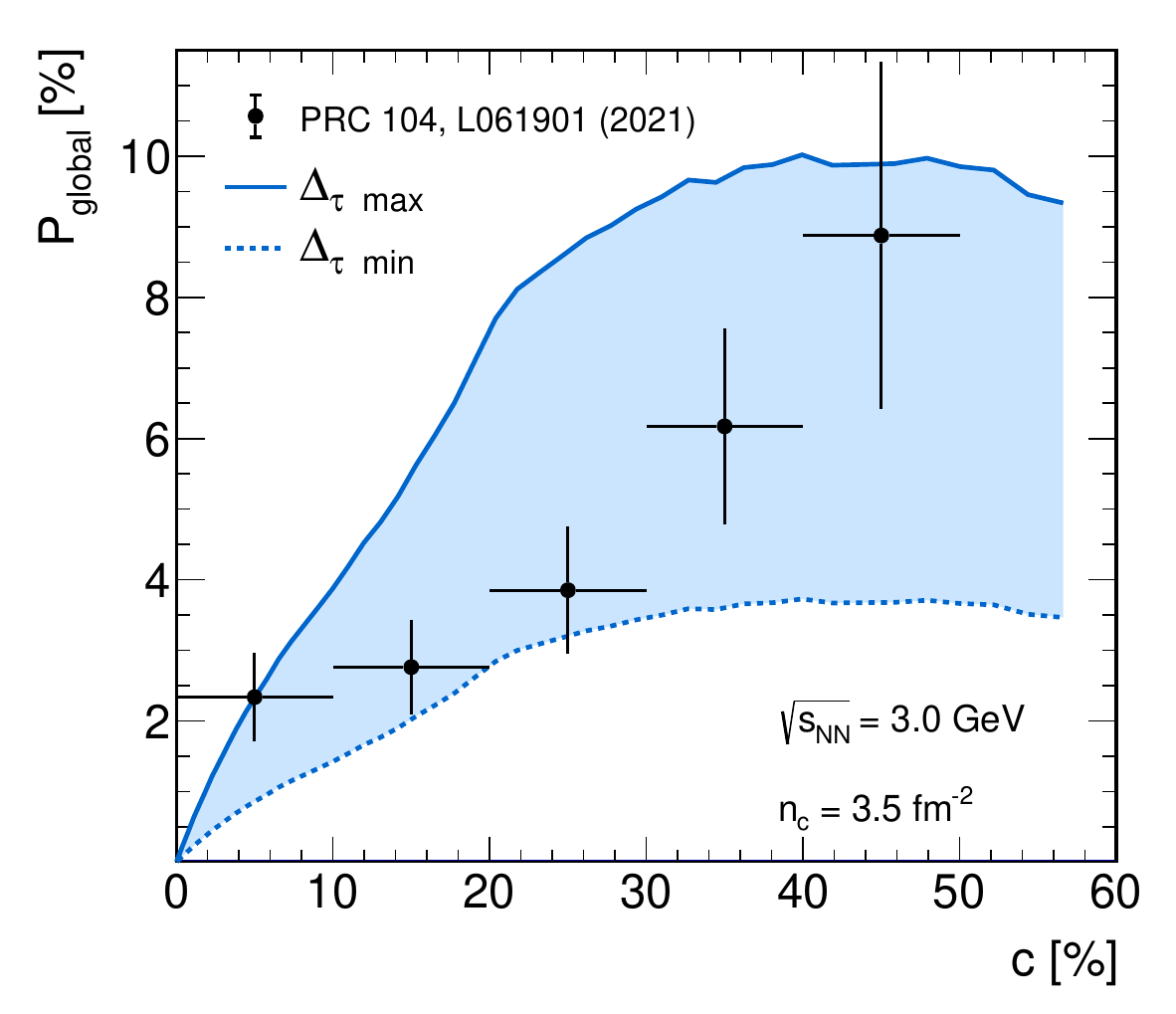}
    \caption{Global $\Lambda$ polarization as a function of centrality for Au+Au $\sqrt{s_{NN}}$ = 3 GeV. The shaded region represents the calculation for life-times between the minimum (lower curve) and maximum (upper curve) life-time estimates of the combined core and corona regions. The results are compared with experimental data from STAR~\cite{STAR:2021beb}.}
    \label{fig:pol_b_3}
\end{figure}
\begin{figure}[b]
    \centering
    \includegraphics[width=\linewidth]{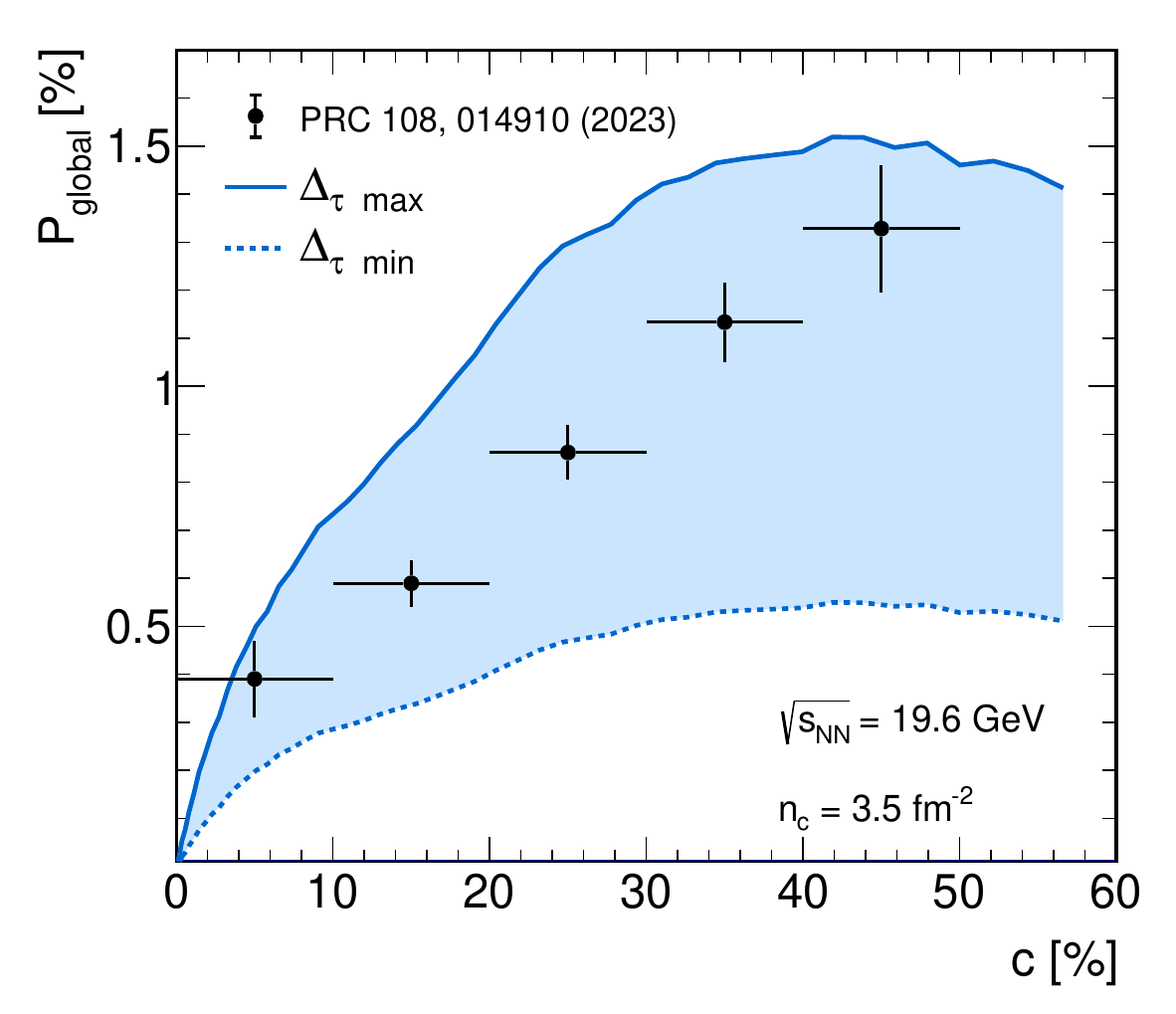}
    \caption{Global $\Lambda$ polarization as function of centrality for Au+Au $\sqrt{s_{NN}}$ = 19.6 GeV. The shaded region represents the calculation for life-times between the minimum (lower curve) and maximum (upper curve) life-time estimates of the combined core and corona regions. The results are compared with experimental data from STAR~\cite{STAR:2023nvo}.}
    \label{fig:pol_b_196}
\end{figure}
\begin{figure}[t]
    \centering
    \includegraphics[width=\linewidth]{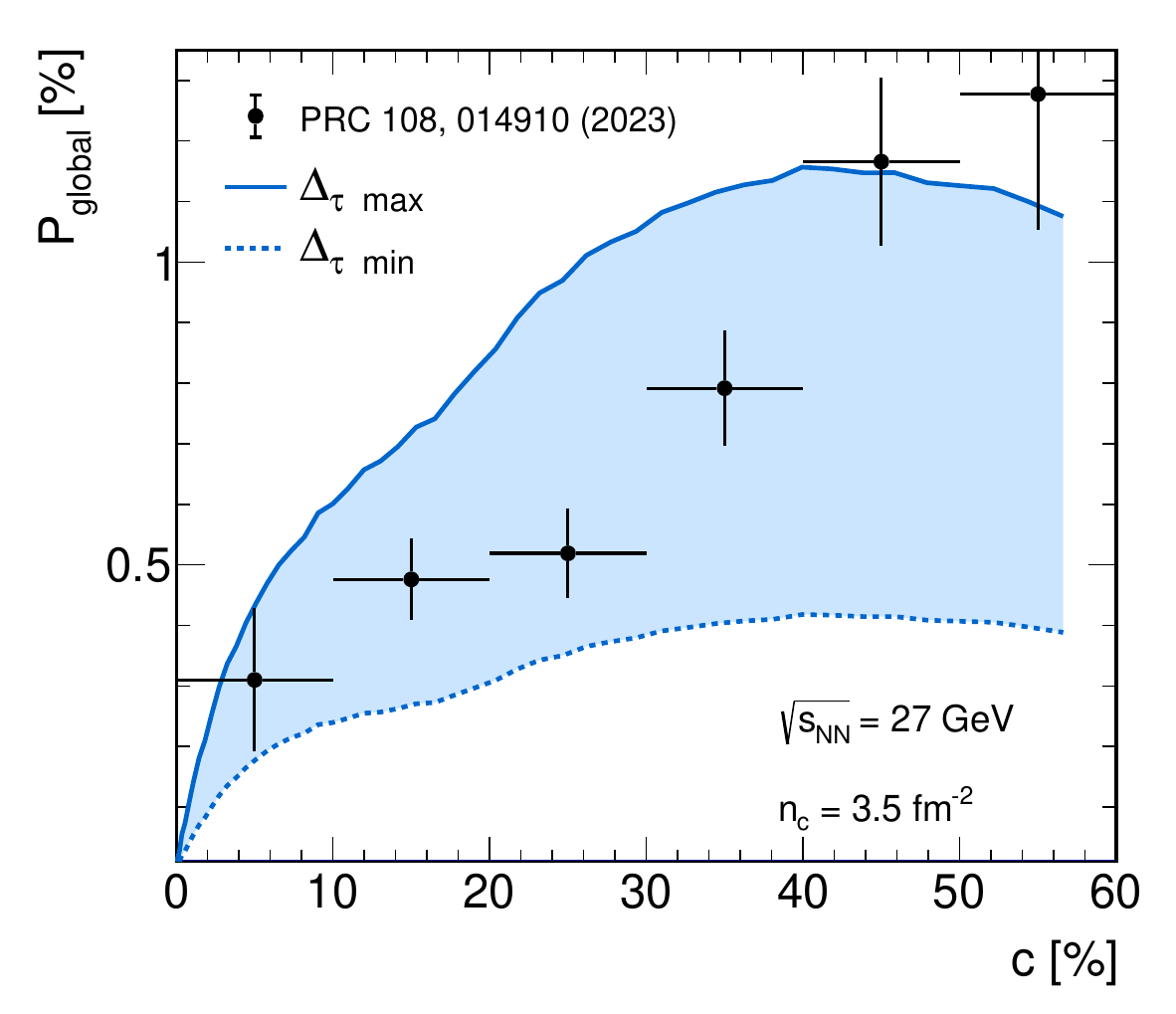}
    \caption{Global $\Lambda$ polarization as function of centrality for Au+Au $\sqrt{s_{NN}}$ = 27 GeV. The shaded region represents the calculation for life-times between the minimum (lower curve) and maximum (upper curve) life-time estimates of the combined core and corona regions. The results are compared with experimental data from STAR~\cite{STAR:2023nvo}.}
    \label{fig:pol_b_27}
\end{figure}
\begin{figure}[b]
    \centering
    \includegraphics[width=\linewidth]{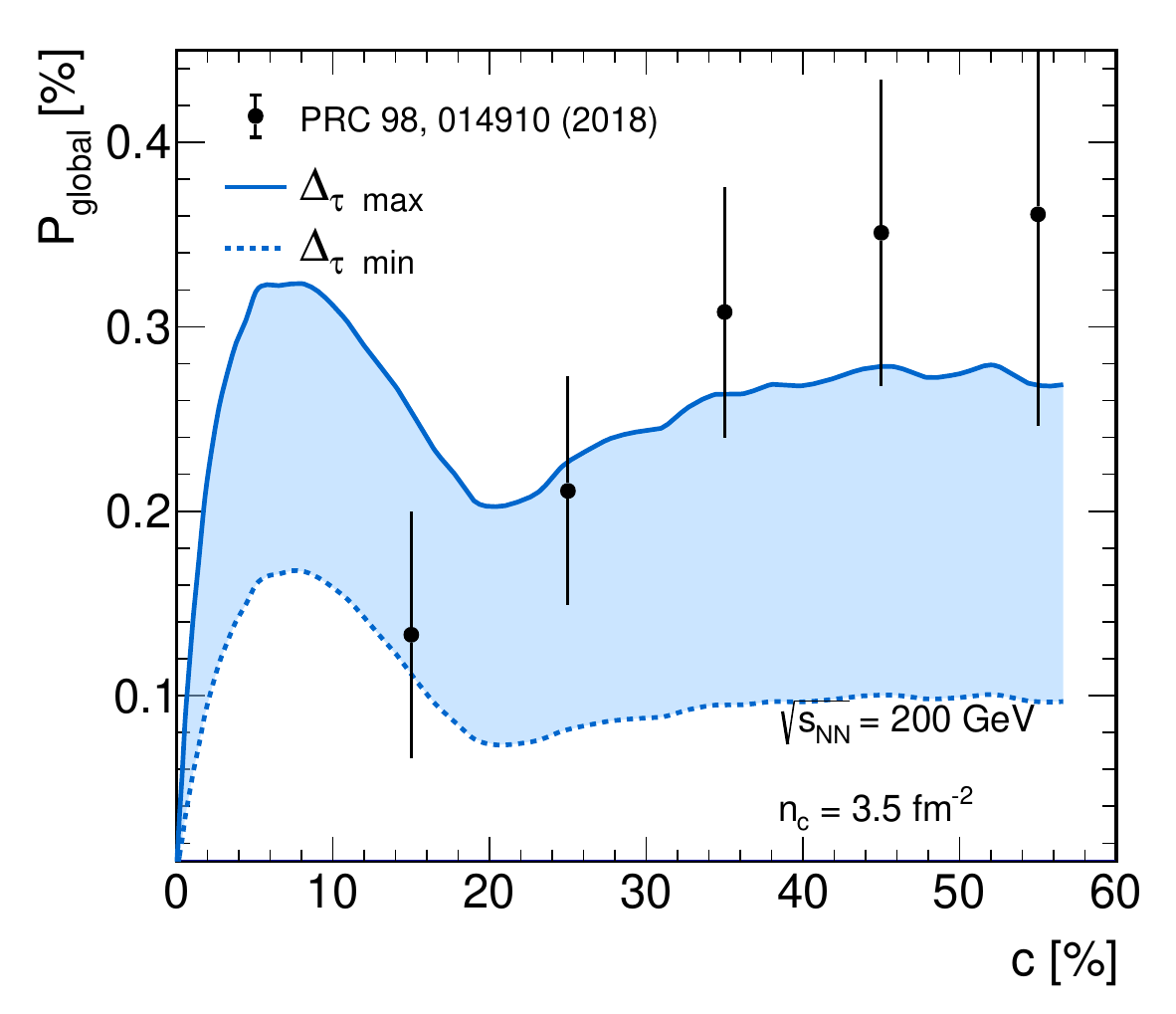}
    \caption{Global $\Lambda$ polarization as function of centrality for Au+Au $\sqrt{s_{NN}}$ = 200 GeV. The shaded region represents the calculation for life-times between the minimum (lower curve) and maximum (upper curve) life-time estimates of the combined core and corona regions. The results are compared with experimental data from STAR~\cite{STAR:2018gyt}.}
    \label{fig:pol_b_200}
\end{figure}
where $\sigma_{NN}^\Lambda$ is the $N+N$ cross-section for the production of $\Lambda$s. For $\sigma_{NN}$ and $\sigma_{NN}^\Lambda$ we use the fits to data reported in Fig. 3 of Refs.~\cite{Ayala:2021xrn,Ayala:2023xyn}. To allow for subthreshold production in a nuclear environment~\cite{HADES:2022enx}, we assume that the effective threshold for $\sigma_{NN}^{\Lambda}$ is shifted down to $\sqrt{s_{NN}}=2.1~\mathrm{GeV}$; below this energy, we set $\sigma_{NN}^{\Lambda}=0$.

Figure~\ref{fig:Nlambda} shows the number of $\Lambda$s produced in the core and in the corona. Notice that this represents an update on the number of $\Lambda$s reported in Ref.~\cite{Ayala:2021xrn} that considers a wider range of collision energies. Figure~\ref{fig:coreregion} illustrates the core and corona regions produced in the transverse plane of the collision for $\sqrt{s_{NN}}=200$ GeV and two impact parameters, $b=0$ (top) and $b=6.66$ fm (bottom) computed using the Glauber model, Eqs.~(\ref{GM}) and~(\ref{thickness}). For a given impact parameter, we compute the transverse areas associated with the core and corona and estimate the corresponding volumes as
\begin{eqnarray}
V_{REC}&=&A_{REC}\Delta\tau_{REC},\nonumber\\
V_{QGP}&=&A_{QGP}\Delta\tau_{QGP},
\label{volumes}
\end{eqnarray}
thereby relating the effective volumes to the corresponding lifetimes.

\section{Global \texorpdfstring{$\Lambda$}{Lambda} polarization}\label{secVI}

To compute the polarization of $\Lambda$ hyperons along the global angular
momentum vector, we use the following parametrization relating the baryon
chemical potential and the temperature along the chemical freeze-out
line~\cite{Andronic:2017pug,Steinbrecher:2018phh}. 
\begin{eqnarray}
    \mu_B(\sqrt{s_{NN}}) &=& \frac{d}{1 + e\sqrt{s_{NN}}},\nonumber\\
    \frac{T(\mu_B)}{T_0} &=& 1 - \kappa_2\left(\frac{\mu_B}{T_0}\right)^2 - \kappa_4 \left(\frac{\mu_B}{T_0}\right)^4,
    \label{freezeout}
\end{eqnarray}
with $d=1.3075$ GeV, $e=$ 0.288 $\text{GeV}^{-1}$, $T_0=$ 0.156 GeV, $\kappa_2=$ 0.0120 and $\kappa_4=$ 0.0025. This parametrization is based on statistical hadronization fits to particle-yield data over a wide range of collision energies and provides a smooth and widely used description of the thermodynamic conditions at freeze-out. In particular, it reproduces the systematic energy dependence of $(\mu_B,T)$ in the beam-energy-scan region
relevant for the present study. Although alternative freeze-out curves exist in
the literature, the differences among them in the energy range considered here
lead only to moderate quantitative variations and do not modify the qualitative
behavior of the polarization excitation function obtained in this work. 

For the angular velocity, we use the parametrization obtained in Ref.~\cite{Ayala:2020ndx}
\begin{eqnarray}
    \Omega(\sqrt{s_{NN}},b) &=& \frac{b^2}{2 V_N}\left[1+ 2 \left( \frac{M_P}{\sqrt{s_{NN}}}  \right)^\frac12 \right],
\end{eqnarray} 
where $V_N = \frac43 \pi R^3$ and $R=1.1$A$^{1/3}$ and $M_P=$ 0.938 GeV is the proton mass. 

The last ingredients for the computation of the polarization are the core and corona lifetimes. Given the uncertainty on these quantities, we resort to set minimum and maximum lifetimes. For the core lifetime, we use the fit to data reported in Fig. 7 of Ref.~\cite{Ayala:2021xrn}
\begin{eqnarray}
    \Delta^{\text{max}}_{\tau\;QGP} &=& 1.166\ln(\sqrt{s_{NN}}/1 {\mbox{GeV}}),\nonumber\\
    \Delta^{\text{min}}_{\tau\;QGP} &=& 0.6803\ln(\sqrt{s_{NN}}/1 {\mbox{GeV}}).
    \label{eq:timeqgp}
\end{eqnarray}
For the corona, we use a similar parametrization, inspired by the findings of Ref.~\cite{Rapp:2014hha,Galatyuk:2015pkq,Kasza:2018qah}

\begin{eqnarray}
   \Delta^{\text{max}}_{\tau\;REC} &\!\!\!=\!\!\!& \left( 2.05\ln(\sqrt{s_{NN}}/1\ {\mbox{GeV}}) + \frac{12.81}{\sqrt{s_{NN}}} - 4.5\right)\; \text{fm}\nonumber\\
    \Delta^{\text{min}}_{\tau\;REC} &\!\!\!=\!\!\!&  \left( 1.23\ln(\sqrt{s_{NN}}/1\ {\mbox{GeV}}) + \frac{7.68}{\sqrt{s_{NN}}} - 2.7\right)\; \text{fm},
    \nonumber\\
    \label{eq:timerec}
\end{eqnarray}

Figure~(\ref{fig:lambdaVol}) shows the core and corona volume dependence with the collision energy $\sqrt{s_{NN}}$ for impact parameters $b = 0, 2, 4$ fm, which is calculated using Eq.~(\ref{volumes}) with the parametrization for the lifetime of Eqs.~(\ref{eq:timeqgp}) and~(\ref{eq:timerec}). Notice that this parametrization accounts for the fact that, due to stopping, the core volume increases at low energies~\cite{Rapp:2014hha,Galatyuk:2015pkq,Kasza:2018qah}.

\begin{figure}[t]
    \centering
    \includegraphics[width=\linewidth]{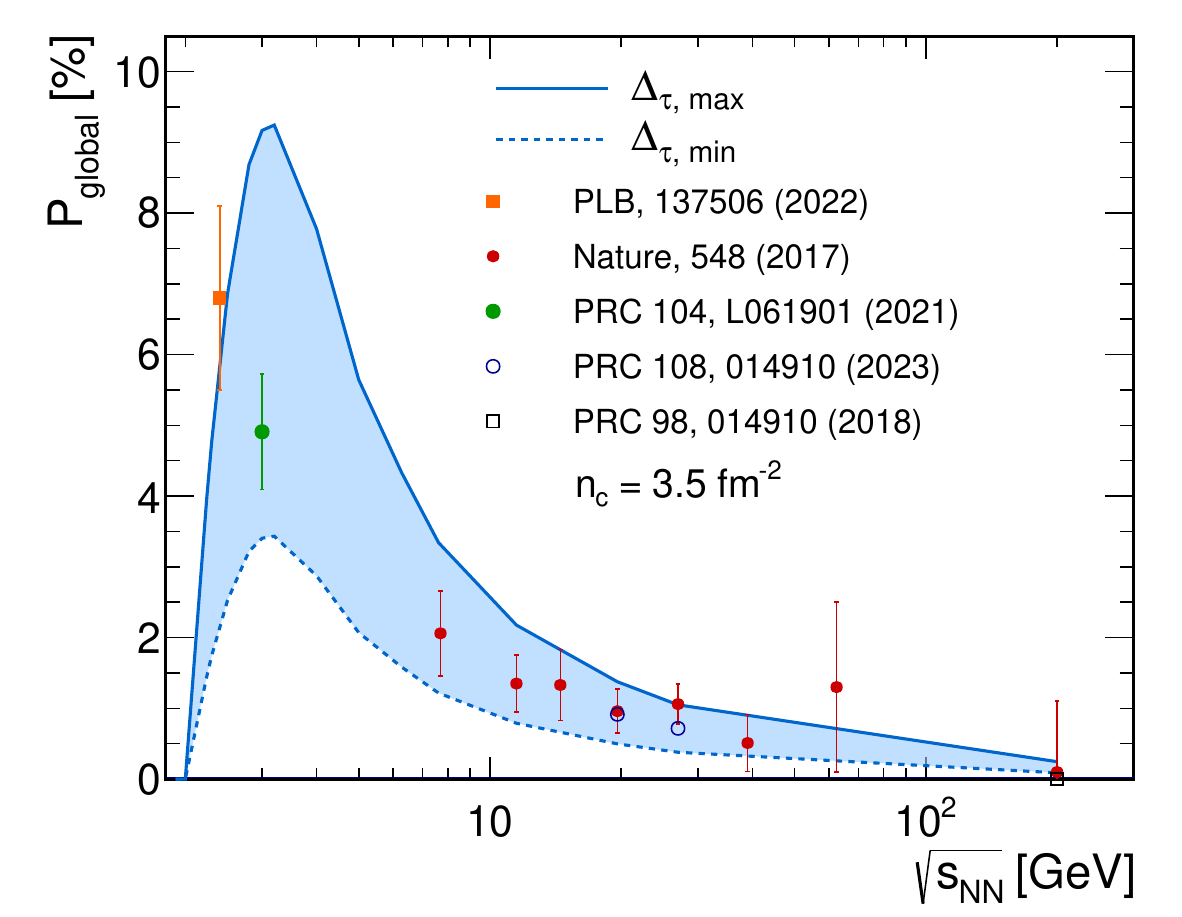}
    \caption{Global $\Lambda$ polarization as a function $\sqrt{s_{NN}}$ compared with experimental data from STAR and HADES~\cite{HADES:2022enx,STAR:2017ckg,STAR:2018gyt,STAR:2021beb,STAR:2023nvo}.}
    \label{fig:polGLOB}
\end{figure}
Figure~\ref{fig:pol_b_3} shows the global $\Lambda$ polarization as a function of centrality for Au+Au collisions at $\sqrt{s_{NN}}$. The boundaries of the shaded region represent the minimum (lower curve) and maximum (upper curve) life-time estimates of the combined core and corona regions. The comparison is made with experimental data from STAR for Au+Au at $\sqrt{s_{NN}}=3$ GeV reported in Ref.~\cite{STAR:2021beb}. Figure~\ref{fig:pol_b_196} shows the global  polarization as a function of centrality for $\sqrt{s_{NN}}$ = 19.6 GeV. Once again, the boundaries of the shaded region represent the minimum (lower curve) and maximum (upper curve) life-time estimates of the combined core and corona regions. The results are compared with experimental data from STAR reported in Ref.~\cite{STAR:2023nvo}. Figure~\ref{fig:pol_b_27} shows the results for the global $\Lambda$ polarization as a function of centrality for Au+Au at $\sqrt{s_{{NN}}}$ = 27 GeV, compared with experimental data from STAR, also reported in Ref.~\cite{STAR:2023nvo}. Figure~\ref{fig:pol_b_200} shows the result for the global $\Lambda$ polarization for $\sqrt{s_{NN}}$ = 200 GeV as a function of centrality, compared with experimental data from STAR reported in Ref.~\cite{STAR:2023nvo}. Notice that in all cases, the experimental data lie, considering the uncertainties,  within the shaded regions and that their rising trend with centrality is well reproduced by the model.

To compute the global $\Lambda$ polarization excitation function that can be compared with experimental data, we need to consider the range of impact parameters corresponding to the reported data centrality range. Therefore, for a fixed energy, we define the average polarization in an impact parameter range as 
\begin{eqnarray}
    P = \frac{\int_{b_{\text{max}}}^{b_\text{min}}P(b)\;db}{b_{\text{max}}-b_{\text{min}}},
\end{eqnarray}
and use this average polarization to compare with the experimental data in the given centrality range. For the range corresponding to the experimental data, namely 20-50\% centrality, $b_\text{min}\approx$ 6.66 fm and $b_\text{max}\approx$ 10.12 fm. Figure~\ref{fig:polGLOB} shows our result of the global polarization for Au+Au collisions as a function of $\sqrt{s_{NN}}$ compared with experimental data from Refs.~\cite{STAR:2017ckg,STAR:2018gyt,STAR:2021beb,STAR:2023nvo}. For completeness, we also show the HADES measurement below the strangeness
production threshold in nucleon-nucleon collisions for Ag+Ag collisions at $\sqrt{s_{NN}}=2.55$ GeV~\cite{HADES:2022enx} in the centrality range 0-40\%. The computed global polarization reaches a maximum near $\sqrt{s_{NN}}\sim 3$ GeV. We have tested the sensitivity of this maximum to parameter variations and found that it can shift toward lower energies as the freeze-out chemical potential $\mu_B$ increases. In the present work, however, we adopt the conservative choice of evaluating the polarization along the freeze-out curve given in Eq.~(\ref{freezeout}). We emphasize that the present calculation relies on two phenomenological inputs that introduce a moderate level of systematic uncertainty, namely the parametrization of the chemical freeze-out curve and the modeling of subthreshold $\Lambda$ production in the nuclear environment. As previously mentioned, different statistical-hadronization analyses lead to
slightly different freeze-out trajectories in the $(T,\mu_B)$ plane; however,
these variations produce only moderate quantitative changes in the polarization
excitation function and do not modify its overall qualitative behavior. Likewise,
the effective treatment of subthreshold $\Lambda$ production through an
energy-dependent nucleon--nucleon production cross section with a lower than usual threshold affects primarily the lowest-energy region of the excitation function. We have verified that reasonable
variations of these inputs do not alter the position of the predicted maximum near
$\sqrt{s_{NN}}\sim 3~\mathrm{GeV}$, which remains a robust feature of the model.

\section{Summary and conclusions}\label{concl}

We have computed the excitation function for the global $\Lambda$ polarization using the Core-Corona model. The model accounts for the contributions to $\Lambda$ production from a high-density region (the core) and a less dense region (the corona) in the nuclear collision. For the former, we model the $\Lambda$ production from QGP processes, whereas for the latter it comes from hadron processes. An important element of the model is the relaxation times for $\Lambda$s to align their spin with the vorticity produced in semicentral collisions in both regions. To compute these relaxation times, we have developed a field theoretical description based on two ingredients: a fermion propagator that carries the information of the medium vortical motion with a constant angular velocity, and suitable mediator propagators at finite density and temperature. In the QGP case, the mediator is the well-known gluon propagator at finite temperature and density in the HTL approximation, whereas for the description of interactions in the corona, we have used the $\sigma$-propagator recently found in Ref.~\cite{Ayala:2025bzk}. The model requires knowledge of the effective volumes and lifetimes of the core and corona. The core volume is modeled as a monotonically increasing function of the collision energy, whereas the corona volume incorporates the increasing relevance of nuclear stopping at low energies. Given the uncertainties of the core and corona lifetimes, we have parametrized them to allow for a lifetime span of about 1 -- 3 fm between the minimum and maximum lifetimes, depending on the collision energy. $\Lambda$ production in the core and the corona is described in terms of a Glauber model, introducing a critical density of participants $n_c=3.5$ fm $^{-2}$ above which QGP is produced, which defines the core region, and below which the density is small, thus defining the region considered as the corona. To account for the subthreshold $\Lambda$ production reported in Ref.~\cite{HADES:2022enx}, we use a $\sigma_{NN}^\Lambda$ energy-dependent cross section that becomes nonzero at $\sqrt{s_{NN}}=2.1$ GeV. With these elements, one can also compute the number of $\Lambda$s coming from the core and the corona, which is another important ingredient of the model. For the centralities where the polarization is experimentally reported, we find that the polarization originates mainly in the corona. The description of the $\Lambda$ polarization excitation function is quite good, all over the range from $\sqrt{s_{NN}}=2.55$ to 200 GeV. We find a maximum of this excitation function at $\sqrt{s_{NN}}\sim$ 3 GeV and a rapid fall for smaller energies down to the considered threshold energy for $\Lambda$ production. The peak position is mainly affected by $\mu_B$ which we have taken from the chemical freeze-out curve reported in Refs.~\cite{Andronic:2017pug,Steinbrecher:2018phh} that determines the relation between the collision energy and the freeze out temperature, and baryon chemical potential. We conclude that, under a set of simple but physically motivated assumptions, the model provides a good description of the $\Lambda$ polarization excitation function. In particular, previous shortcomings of the model that prevented it from describing the HADES point are fixed simply by allowing a lower threshold for $\Lambda$ production in the nuclear environment compared to proton-proton collisions. We also emphasize that, for the model to describe the data,  it is important to account for a larger lifetime and, consequently, a larger volume of the corona for energies smaller than $\sqrt{s_{NN}}\sim 7$ GeV. Finally, our prediction of the rising trend for smaller energies up to a maximum, followed by a rapid fall, seems to be a robust feature of the model.

\section*{Acknowledgements}
A.A. thanks the colleagues and staff of Universidade de São Paulo, of Instituto de F\'isica Te\'orica, UNESP and of Universidade Cidade de São Paulo for their kind hospitality during a sabbatical stay in which part of this work was carried out. A.A. also acknowledges support from the PASPA program of Direcci\'on General de Asuntos del Personal Acad\'emico (DGAPA) of the Universidad Nacional Aut\'onoma de M\'exico (UNAM) for the sabbatical stay during which this research was carried out and for support via grant number IG100826. Support for this work has been received in part via Secretar\'\i a de Ciencia, Humanidades, Tecnolog\'\i a e Innovaci\'on (SECIHTI) M\'exico grant number CIORGANISMOS-2025-17.

\bibliography{bibliosigma}

\end{document}